\documentclass{ametsocV6.1}

\title{Comparison of Two Operational Microphysics Schemes Across Various Regional-MPAS Simulations}

\authors{
Abraham I. Roseman\aff{1},
Falko Judt\aff{2},
Wei Wang\aff{2},
Ting-Yu Cha\aff{2},
Giuseppe Torri\aff{1}
}

\usepackage[utf8]{inputenc}
\usepackage{textcomp}
\DeclareUnicodeCharacter{02BB}{\okina}
\newcommand{\okina}{\textquoteleft}

\affiliation{
\aff{1}{Department of Atmospheric Sciences, University of Hawai\okina i at M\=anoa, Honolulu, Hawai\okina i, USA}\\
\aff{2}{Mesoscale \& Microscale Meteorology Laboratory, National Center for Atmospheric Research, Boulder, Colorado, USA}
}

\abstract{Accurately representing convection and precipitation remains a persistent challenge for Numerical Weather Prediction (NWP) models due to biases in convective initiation, storm organization, and rainfall distribution, particularly in subtropical/tropical environments. This study evaluated how bulk microphysics parameterizations influence convective organization and precipitation using convection-permitting hindcasts with the Model for Prediction Across Scales - Atmosphere (MPAS-A) on a variable-resolution mesh down to 1-km resolution. Two operational microphysics schemes, National Severe Storm Labs (NSSL) microphysics and Thompson-Eidhammer Microphysics Parameterization for Operations (TEMPO), were examined across three subtropical/tropical regions during boreal summer under strongly- and weakly-forced regimes. Both schemes captured the general timing and large-scale placement of convection, but differed in storm structure and rainfall distribution. TEMPO produced more numerous, weaker convective cores with earlier, more widespread precipitation and cooler surface conditions, while NSSL favored fewer, stronger cores and updrafts with more cloud water, ice, and graupel hydrometeors, though less snow, and more spatially concentrated, intense rainfall. Despite these structural differences, both schemes diverged more from observations than from each other, producing scattered convective cells with minimal mesoscale organization and insufficient stratiform precipitation. The simulations also exhibited regime-dependent errors, with rainfall under- (over)-represented in strongly- (weakly)-forced regimes and forecast skill notably lower in the latter. Improving the representation of localized precipitation processes remains essential for capturing convection across a wider range of scales and regimes. Future work should include targeted evaluation of microphysics across regimes and regions, along with process-level improvements to address convective biases.}

\begin{document}

\maketitle

%
%
%
%
%
%

\statement{Reliable forecasts of convection and precipitation are essential for local weather prediction, particularly in subtropical and tropical regions, where forecast errors remain large. This study shows that the choice of microphysics scheme, which represents how cloud particles form and become precipitation in a weather forecast model, can strongly affect storm development. Two operational microphysics setups, NSSL and TEMPO, were compared for strongly- and weakly-forced weather regimes, across multiple regions. TEMPO produced broader, weaker convection, while NSSL produced stronger, more localized storms. These findings highlight key considerations for forecasters when selecting microphysics setups. However, both schemes shared larger biases relative to observations, including overly scattered convection and weak mesoscale organization, which require further improvement to support reliable storm-scale forecasts.}

%








\section{Introduction}

Modern Numerical Weather Prediction (NWP) can reproduce large- and local-scale atmospheric dynamics with considerable skill at convective-resolving resolutions, with global 3-km simulations feasible on large supercomputers \citep{Heinzeller2016}. Nevertheless, deficiencies in convective-scale predictably remain large \citep{Yano2018}. Deep convection drives many high-impact, hazardous weather hazards, yet its representation in regional forecasts remains limited. Improving convective-scale prediction is therefore essential for providing earlier and more reliable guidance for local communities, with broad societal benefits for disaster preparedness, agriculture, and water resource management. Addressing deficiencies in convective-process representation remains a key step toward improving NWP.

Common convective deficiencies include errors in initiation timing, spatial organization, intensity, and structure, which can propagate into larger-scale forecast errors. Previous convective-allowing simulations have shown substantial errors in convective initiation (CI), including timing errors of several hours and spatial displacement errors exceeding 100 km \citep{Duda2013,Stelten2017}, while higher-resolution simulations can reduce some displacement errors, but increase false alarms due to excessive CI \citep{Burghardt2014}. These errors often manifest as precipitation biases, where convective-permitting systems, have shown both overestimation and underprediction depending on region, regime, and model microphysics configuration \citep{Fowler2016,Fowler2020,Heinzeller2016,Guo2019,Johnson2023,deSouza2023}.

One approach that reduces convective biases is to increase horizontal resolution to spacings at kilometer scale or lower to improve representation of small-scale forcing related to CI and individual storm evolution \citep{Duda2013}. Additionally, precipitation biases, such as overly drizzle-like precipitation, has been reduced by explicitly resolving convection \citep{Cheng2026}. Recent advances in computing have enabled kilometer-scale convective-allowing regional and global prediction systems, including the High-Resolution Rapid Refresh (HRRR) \citep{Dowell2022}, the Next Generation Global Prediction System (NGGPS) \citep{Zhou2019}, and NOAA’s 3-km MPAS-based operational Unified Forecast System (UFS) \citep{aligo2026}. Improvements in variable-resolution nested meshes introduced by MPAS-A help overcome kilometer-resolution computational constraints by enabling localized refinement within coarser domains. The scale-aware New Tiedtke scheme \citep{Wang2022} was developed to reduce mesh-dependent convective sensitivity biases identified in previous studies \citep{Fowler2016,Fowler2020}, including excessive localized rainfall. This allows convective parameterizations to represent unresolved convection in coarser regions, while at higher resolutions, where clouds become partially resolved, explicit microphysics plays a larger role in representing convective structure.

\cite{Feng2018} demonstrated that microphysics schemes strongly modulate both convective dynamics and storm properties. As a result, convective forecasts remain highly sensitive to microphysics choice, particularly at kilometer-scale resolutions and finer. This current study examines how microphysics parameterization modulates convection and precipitation in 1-km convection-permitting simulations and evaluates associated deficiencies across schemes within Model for Prediction Across Scales - Atmosphere (MPAS-A). This study compares National Severe Storm Labs (NSSL) microphysics \citep{Mansell2010} and Thompson-Eidhammer Microphysics Parameterization for Operations (TEMPO) \citep{Thompson2004,Thompson2008,Thompson2014,Jensen_A_2023,Jensen2024} schemes across seasonally-active subtropical/tropical regimes. The focus on subtropical and tropical regimes is motivated by well-documented convection biases in these regions \citep{Maithel2026}, including too early precipitation, excessive precipitation, and insufficient moistening or artificial drying \citep{Hirons2013,Wolding2020,Kuo2020,Han2021,Ye2023}. Because tropical convection varies strongly across diverse tropical environments, many cloud and precipitation processes occur at scales too fine for standard global models to resolve explicitly, requiring parameterization and complicating accurate tropical precipitation forecasts \citep{Houze2015}. These tropical latent heating biases can also propagate to higher latitudes \citep{Dias2019}, thus are more broadly important for forecasting globally.

TEMPO is an operational version of the Thompson-Eidhammer scheme, for use in operational forecast systems. Thompson-Eidhammer is two-moment in cloud water, cloud ice, and rain (generalized-gamma distribution), and one-moment in all others (exponential distribution) \citep{Iversen2021}, while TEMPO introduces hail- and aerosol-aware settings, the former introducing two-moment variable density graupel. On the other hand, NSSL is a fully two-moment scheme employing generalized gamma size distributions for all hydrometeors. NSSL also includes more detailed ice microphysics (e.g. secondary ice processes). Cloud droplet number concentration is also a constant value in TEMPO, while prescribed in NSSL. \cite{Johnson2023} found that both NSSL and Thompson underpredicted overall storm coverage, but NSSL overpredicted high-reflectivity convective coverage, while Thompson underpredicted it. Thompson has also been shown to favor snow over other ice species relative to other schemes \citep{Guo2019}, though this behavior may not extend to NSSL. As a later version of the Thompson scheme, prior results for Thompson provide context, though they may not transfer directly. NSSL’s fully two-moment, generalized-gamma formulation may improve convective structure and rainfall representation relative to TEMPO at kilometer scales, consistent with the benefits of two-moment schemes noted by \cite{Bryan2012}. Similarly, \cite{chen2026} found that NSSL improved precipitation skill relative to Thompson in a convection-permitting MPAS system over the Maritime Continent.

Together, persistent convective prediction biases and microphysics sensitivities motivate a focused evaluation of microphysics in Regional-MPAS. This evaluation can help guide microphysics selection in emerging operational Regional-MPAS forecasting systems. Section 2 describes the methods, including case selection, model simulations, and data processing. Section 3 presents the results, comparing simulated precipitation and radar reflectivity with observations, as well as microphysical fields between schemes. Section 4 provides an overall summary and discussion of the findings. Section 5 concludes with key findings and recommended future work.

\section{Methods}
\subsection{Case Descriptions}

High-resolution, multi-day Regional-MPAS hindcasts were conducted over Houston/Galveston, Taiwan, and Hawai\okina i during the boreal summer. Six cases were selected to represent strongly-forced (“WET”) and weakly-forced (“DIURNAL”) convective regimes, with the Hawai\okina i trade-wind case classified separately as “TRADES.” WET cases featured widespread, organized convection; DIURNAL cases were dominated by thermally-driven coastal convection; and the TRADES case represented shallow cumulus and terrain-enhanced rainfall over windward and mountainous inland (mauka) regions. A “DRY” regime was excluded because both schemes produced widespread spurious convection, limiting meaningful comparison with observations. For each case, separate simulations were conducted using NSSL and TEMPO microphysics. Various observations and resources were used to identify cases and CI timing. Observations from the 2021-2022 Tracking Aerosol Convection Interactions Experiment (TRACER) campaign \citep{Jensen_M_2023} was used for Houston, 2022 National Science Foundation (NSF)-supported Prediction of Rainfall Extremes Campaign in the Pacific (PRECIP) experiment \citep{Zhang2023} for Taiwan, and Hawai\okina i observations were sourced from local operational datasets. Simulations were initialized 12 hours before observed CI, which best captured convective timing and placement compared with 6-hour and 24-hour spinup in sensitivity tests. For Houston and Hawai\okina i, Multi-Radar Multi-Sensor (MRMS) mosaic reflectivity \citep{Zhang2011,Zhang2016} observations were used. Additional TRACER observations included Ka- and X-band radar, radiosondes, and surface meteorological measurements from the ARM Mobile Facility One \citep{Hardin2018,Isom2018,Keeler1994,Miller2016}. S-Pol radar (Hsinchu, Taiwan) observations from the PRECIP campaign were used for Taiwan. National Weather Service (NWS) Area Forecast Discussions (AFDs), obtained from the Iowa Environmental Mesonet \citep{Todey2002} and PRECIP daily reports provided context about weather evolution.

The selected cases and their weather characteristics are summarized below and relevant model-observation comparisons are shown in Figure~\ref{figure3} (rainfall) and Figure~\ref{figure6} (horizontal reflectivity). Simulation periods for each case are shown in Table~\ref{table1}. The TRACER/WET case (1 July 2022) featured widespread, organized coastal convection associated with a coastal low and an enhanced Gulf moisture transport with widespread rainfall developing across SE Texas and SW Louisiana with scattered convection persisting as the system shifted NEward. The TRACER/DIURNAL case (22 June 2022) was primarily sea-breeze driven, with isolated afternoon convection forming along the coast and moving inland near Houston before dissipating by evening. The PRECIP/WET case (6 June 2022) was associated with a southward-moving Mei-Yu front that later became quasi-stationary near Taiwan, producing widespread heavy rainfall over northern and central Taiwan. A PRECIP Intense Operating Period (IOP) began at 0600 UTC continuing until 1200 UTC on 12 June 2022 \citep{Yang2024}. Low-level southwesterly flow supported moisture transport and convection along the Mei-Yu front and western coast later merged into a rainband that produced heavy rainfall over NW and N Taiwan. Rainfall rates reached about $90\ \mathrm{mm\,hr^{-1}}$ near the western coast and accumulated rainfall approached $300\ \mathrm{mm}$ within the first three days. The PRECIP/DIURNAL case (16 July 2022) featured isolated, short-lived shallow cells forming near the NW and central coast before moving inland. The Hawai\okina i/WET case (6 December 2021) was associated with a Kona low that developed after frontal passage, as an upper-level low cut off from the polar jet northwest of Hawai\okina i. Cold air aloft interacting with moist tropical flow produced widespread heavy rainfall and gusty winds, including a north-south-oriented rainband over O\okina ahu on the evening of the 6th, producing most of the event's rainfall. Forecast skill for Hawai\okina i/WET was sensitive to initialization due to its multiple rainfall episodes; although a later initialization (0400 UTC 7 December 2022) better captured the peak rainfall event, instead an earlier initialization was used to capture the full case evolution. The Hawai\okina i/TRADES case (7 August 2022) featured a dry trade-wind regime, with a subtropical high maintaining easterly flow, limited windward and mauka showers, and suppressed deep convection under a strong subsidence inversion.

\begin{table}[t]
\caption{Summary of the six Regional-MPAS case studies.}
\label{table1}
\begin{center}
\begin{tabular}{lllll}
\hline\hline
\textbf{Case} & \textbf{Region} & \textbf{Simulation Start} & \textbf{Simulation End} & \textbf{Observed Forcing} \\
\hline
TRACER/WET & Houston/Galveston & 2022/07/01 0000 UTC & 07/03 0000 UTC & Coastal low; widespread organized convection \\
TRACER/DIURNAL & Houston/Galveston & 2022/06/22 0600 UTC & 06/24 0000 UTC & Sea-breeze-driven isolated convection \\
PRECIP/WET & Taiwan & 2022/06/05 1200 UTC & 06/08 0000 UTC & Mei-Yu front; widespread heavy rainfall \\
PRECIP/DIURNAL & Taiwan & 2022/07/15 1200 UTC & 07/18 0000 UTC & Weakly forced coastal/diurnal convection \\
Hawai\okina i/WET & Hawai\okina i & 2021/12/05 1200 UTC & 12/08 0000 UTC & Kona low; widespread heavy rainfall \\
Hawai\okina i/TRADES & Hawai\okina i & 2022/08/07 0000 UTC & 08/10 0000 UTC & Trade-wind showers; shallow terrain-linked precipitation \\
\hline
\end{tabular}
\end{center}
\end{table}

\subsection{Model Description}

\begin{figure*}[t]
 \noindent\includegraphics[width=\textwidth,angle=0,trim=0 9 0 0]{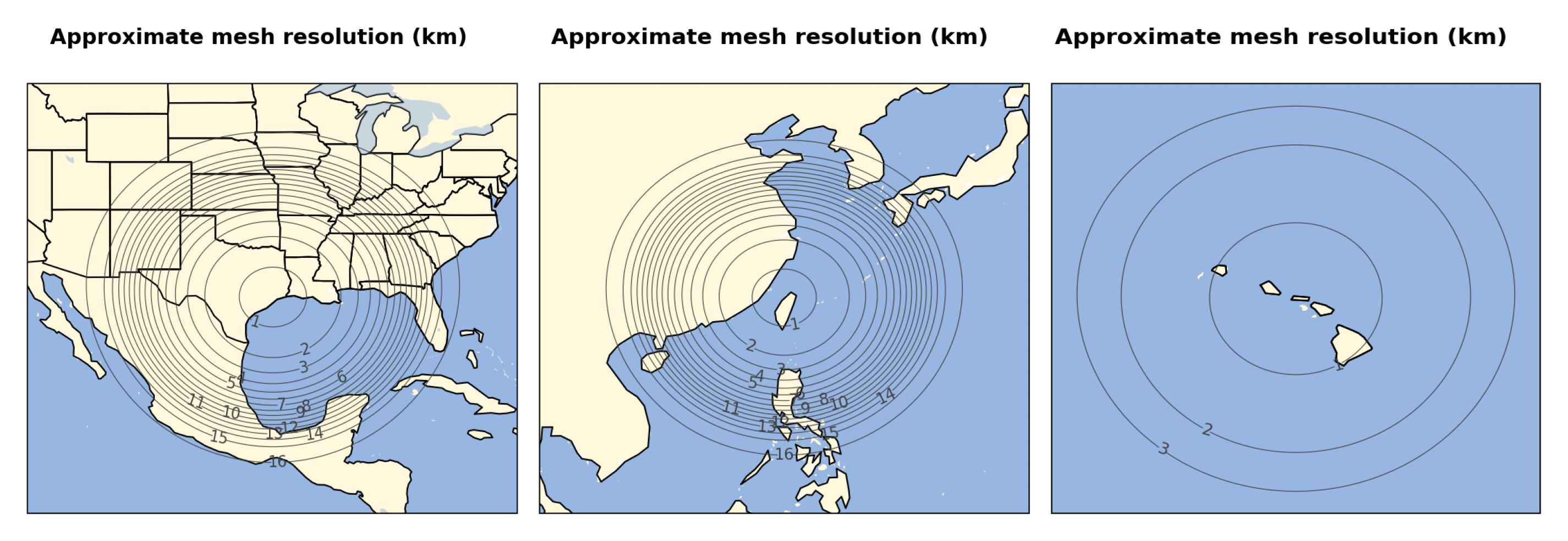}
 \caption{Approximate horizontal mesh resolution in kilometers for the regional MPAS-A domains over Houston (left), Taiwan (center), and Hawai\okina i (right).}\label{figure1}
\end{figure*}

Simulations were conducted using MPAS-A \citep{Skamarock2012}, a fully compressible, non-hydrostatic model with a centroidal Voronoi mesh enabling seamless regional refinement without nesting. Preprocessing used MPAS v8.3.1, with static fields from the NCAR MMM repository, including land use (IGBP-MODIS Noah), soil type (STATSGO), topography (GMTED2010), and MODIS-based vegetation and albedo. Initial (ICs) and lateral boundary (BCs) conditions were generated from ERA5 reanalysis at 0.25$^\circ$ resolution \citep{Hersbach2020} using the \textit{era5\_to\_int} tool \citep{github-Era5_to_int}. Each domain (Figure~\ref{figure1}) was generated from a global 60-3 km variable-resolution mesh (835,586 cells; $\sim\!16^\circ$ refinement region) and converted to a 20-1 km regional mesh using the \textit{MPAS-Limited-Area} and \textit{scale-region} tools \citep{github-mpas_limited_area,github-scale_region}. The 1-km inner mesh permits explicit convection, while the 20-km outer mesh provides a smoother transition to ERA5 forcing. The model used 56 stretched vertical levels up to 30 km, with 15--50 m near-surface spacing increasing to 500--1000 m aloft, remaining constant above $\sim\!15$ km. Simulations were then produced with the GSL version of MPAS v8.3.0 \citep{github-mpas_dev-microted} to access the NSSL and TEMPO microphysics schemes. Default MPAS-A settings were used except for timestep and damping, with a 6-s timestep satisfying CFL stability requirements. Simulations were compiled with Intel compilers and run on NSF NCAR Derecho using 1024 processors across 8 AMD Milan nodes each with 235 GB memory. Wall-clock time was $\sim\!2.5$ hours per forecast day, with TEMPO $\sim\!15\%$ faster than NSSL. Native MPAS output was converted to a 500 km $\times$ 500 km, 0.01$^\circ$ (1-km) structured Cartesian grid using \textit{convert\_mpas} utility \citep{github-scale_region}, using grid bounds defined about the domain center using a local latitude-longitude approximation for accuracy. Latitudinal and longitudinal extents were estimated as $\Delta \mathrm{Lat}=\Delta Y/R$ and $\Delta \mathrm{Lon}=\Delta X/[R\cos(\mathrm{Lat})]$, then applied symmetrically about the domain center and rounded to the nearest degree. Three-dimensional variables were vertically interpolated from terrain-following coordinates to MSL height using each domain’s model MSL height variable.

Significant model-top instabilities formed at model top for the Hawai\okina i/WET case, requiring a correction to damping in the refinement region for this case alone. An additional instability issue in the uppermost 10 km of the model required increased upper-level damping in this region, including a second-order horizontal filter in the gravity-wave absorbing layer (MPAS-CAM; $\mu = 8.333 \,\times \Delta x \,\times \,0.2\ \mathrm{m^2\,s^{-1}}$) and Rayleigh damping on horizontal velocity. These adjustments stabilized the simulation with minimal impact on lower levels, as damping was confined to levels near model top.

All simulations employed consistent physics parameterizations, described in Table~\ref{table2}. Planetary boundary layer (PBL) turbulent mixing was represented using the scale-aware MYNN eddy-diffusivity mass-flux (MYNN-EDMF) scheme, while horizontal turbulent mixing is handled by MPAS-A’s 2D Smagorinsky closure with a mixing length set to the finest mesh spacing (i.e. 1 km). In the model setup, the PBL scheme is not considered in the radiation calculation and thus does not double count with the Xu and Randall cloud fraction scheme. Convection was parameterized using the scale-aware New Tiedtke scheme \citep{Wang2022}, which reduces parameterized convective tendencies as grid spacing decreases, supporting a transition toward explicit convection at kilometer scales from parameterized convection over coarser mesh regions. Shallow convection overlap between MYNN-EDMF and New Tiedtke was small and did not appreciably affect the results in the short simulations analyzed here. This choice follows \cite{deSouza2023}, who found strong MPAS-A performance using Thompson with New Tiedtke in a tropical-transition cyclone case. For both microphysics schemes, the \textit{config\_microp\_re} option was used to couple microphysics-derived effective radii with RRTMG radiation. The hail-aware (aerosol-aware) option was enabled (disabled) in TEMPO for consistency with NSSL.

\begin{table}[t]
\caption{Physics parameterizations used in MPAS-A simulations.}
\label{table2}
\begin{center}
\begin{tabular}{lll}
\hline\hline
\textbf{Physics} & \textbf{Parameterization} & \textbf{Reference} \\
\hline
Radiation & RRTMG & \cite{Iacono2008} \\
Land surface & Noah-MP & \cite{Niu2011} \\
Surface layer & MYNN & \cite{Olson2021} \\
Planetary boundary layer & MYNN-EDMF & \cite{Olson2019} \\
Orographic gravity-wave drag & YSU & \cite{Hong2008} \\
Convection & New Tiedtke & Wang 2022 \\
Cloud Fraction & Xu-Randall & \cite{Xu1996} \\
Microphysics & NSSL / TEMPO & \cite{Mansell2010} / \cite{Jensen_A_2023} \\
\hline
\end{tabular}
\end{center}
\end{table}

\subsection{Data and Analysis Framework}

This section describes the observational datasets and verification methods used to evaluate the model simulations. Large-scale behavior was evaluated by comparing 500-hPa geopotential height and 850-hPa winds with ERA5 data and 2-m temperature from the model grid cell nearest to each available surface observation from TRACER, PRECIP raw QPESUMS data, and Hawai\okina i weather stations. Model precipitation was evaluated using metrics from gridded quantitative precipitation estimation (QPE) products, including hourly, 0.01$^\circ$ MRMS QPE \citep{Smith2016} data for TRACER and Hawai\okina i cases and MRMS QPE and Segregation Using Multiple Sensors (QPESUMS) \citep{Chang2021} hourly data for PRECIP cases. To validate model convection, reflectivity was evaluated against observations selected nearest to model time. MRMS reflectivity (\textit{MergedReflectivityQC}) data available every 2 minutes was used for TRACER and Hawai\okina i, and PRECIP used S-Pol radar reflectivity (\textit{DBZ\_F\_L2}), gridded with the Lidar Radar Open Software Environment (LROSE) \citep{Dixon2026}. Model output was interpolated to the observational grid for comparison. Only reflectivity values $\ge 0$ dBZ were included for both model and observation data to exclude clear-air signals. Because radar-based evaluation depends on observational sampling, scanning geometry, spatial coverage, and model-derived reflectivity, comparisons are interpreted primarily as structural rather than exact pointwise validation. 

Fraction Skill Score (FSS) \citep{Antonio2025}, calculating using the scores Python library \citep{Leeuwenburg2024}, was used to quantify spatial agreement between modeled and observed reflectivity exceedance at selected thresholds within a 5-km neighborhood. Contoured Frequency by Altitude Diagrams (CFADs), normalized by height, were used to evaluate statistical vertical reflectivity structure, using a custom-adapted version of the Py-ART \textit{create\_cfad} function \citep{Helmus2016} for gridded, multi-time data. For PRECIP cases, an additional signal-to-noise (SNR) filter $\text{Noise}_{1\text{km}} + 20 \log_{10}(r)$ was applied where $r$ is the radial distance from the radar to the corresponding model grid location, and $\text{Noise}_{1\text{km}} = -42.4$ is the minimum detectable signal of the S-Pol radar at a radial distance of 1 km. This improves model reflectivity consistency with observations by approximating and applying observation-consistent distance-dependent radar sensitivity quality control (QC) to model data. This approach was not applied to MRMS because internal QC information is unavailable and cannot be replicated in model output; nevertheless, MRMS comparisons remain useful for assessing relative inter-scheme differences. Area-averaged height-time cross sections were then computed for modeled microphysical fields. Horizontal averages were cosine-latitude weighted and limited to grid points within the available observed radar coverage.

\section{Results}
\subsection{Large-scale Model Validation}

\begin{figure*}[t]
 \noindent\includegraphics[width=\textwidth,angle=0,trim=0 9 0 0]{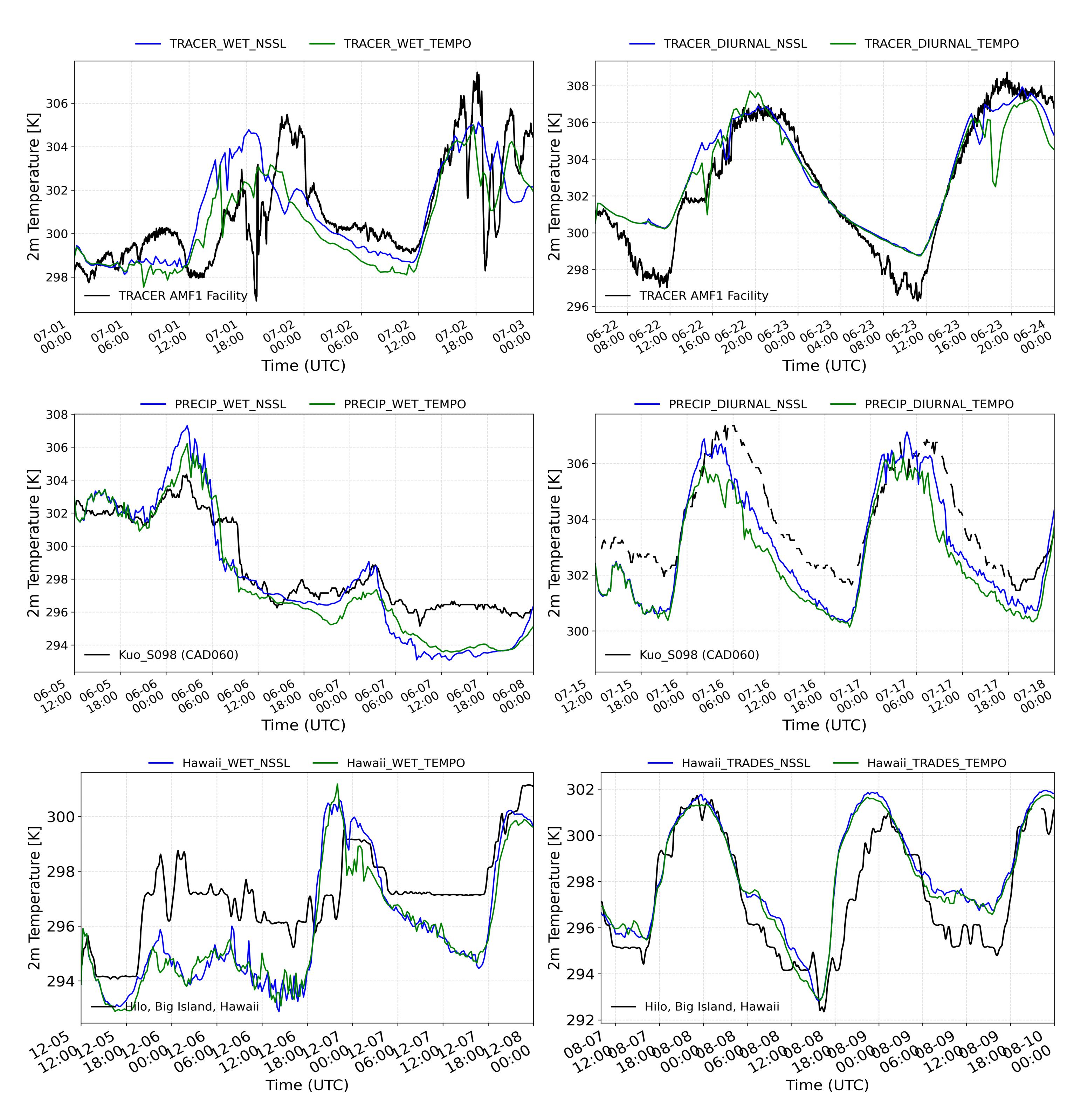}
 \caption{Time series of 2-m air temperature from surface observations (black) and model simulations at the nearest grid point to each station (blue: NSSL; green: TEMPO). Observations are compared with model output from the lowest model level at the corresponding nearest grid point: TRACER ARM1 at 8.14 m MSL compared with model output at 8.0 m MSL; PRECIP/Taiwan QPESUMS CAD110 at 17.0 m MSL versus 10.52 m MSL; and Hawai\okina i MesoWest PHTO at 8.18 m MSL versus 10.97 m MSL. Rows correspond to TRACER, PRECIP/Taiwan, and Hawai\okina i, and columns show WET cases on the left and DIURNAL (TRADES for Hawai\okina i) cases on the right.}\label{figure2}
\end{figure*}

All simulations reproduced the broad spatial and temporal evolution of ERA5 reasonably well (not shown). In TRACER/WET, both schemes captured the initial coastal low and associated dynamics throughout, but failed to clearly maintain the low, which weakened and became smoothed in surface pressure. A recurring bias was weaker smoothed-out geopotential height gradients than ERA5, likely related in part to model spin-up and adjustment from ICs. A recurring issue in ERA5 was coherent circular, dipole, or elongated wind structures that were absent in the simulations and appeared unrealistic at regional scales. Outside of TRACER cases, simulations agreed more closely with ERA5. Across all regimes, simulations captured the diurnal cycle well (Figure~\ref{figure2}), but smaller-scale variability was underrepresented, producing biases of up to $\sim$2 K on average. In TRACER, biases were generally within $\sim$1--2 K, though fine-scale variability remained unresolved; TRACER/DIURNAL showed stronger agreement, except for an underestimated temperature drop between peaks. In PRECIP cases, biases mostly remained within a few tenths of a degree, except for brief $\sim$1--2 K underestimation during later cooling periods. In Hawai\okina i/WET, temperature was underestimated by about 2--4 K early in the simulation, with improved agreement later, but then another underestimation during the O\okina ahu rain event, while Hawai\okina i/TRADES more closely matching the diurnal cycle, with only brief overestimations of 1--2 K.

\subsection{Precipitation and Reflectivity Comparison}

\begin{figure*}[t]
 \noindent\includegraphics[width=\textwidth,angle=0,trim=0 9 0 0]{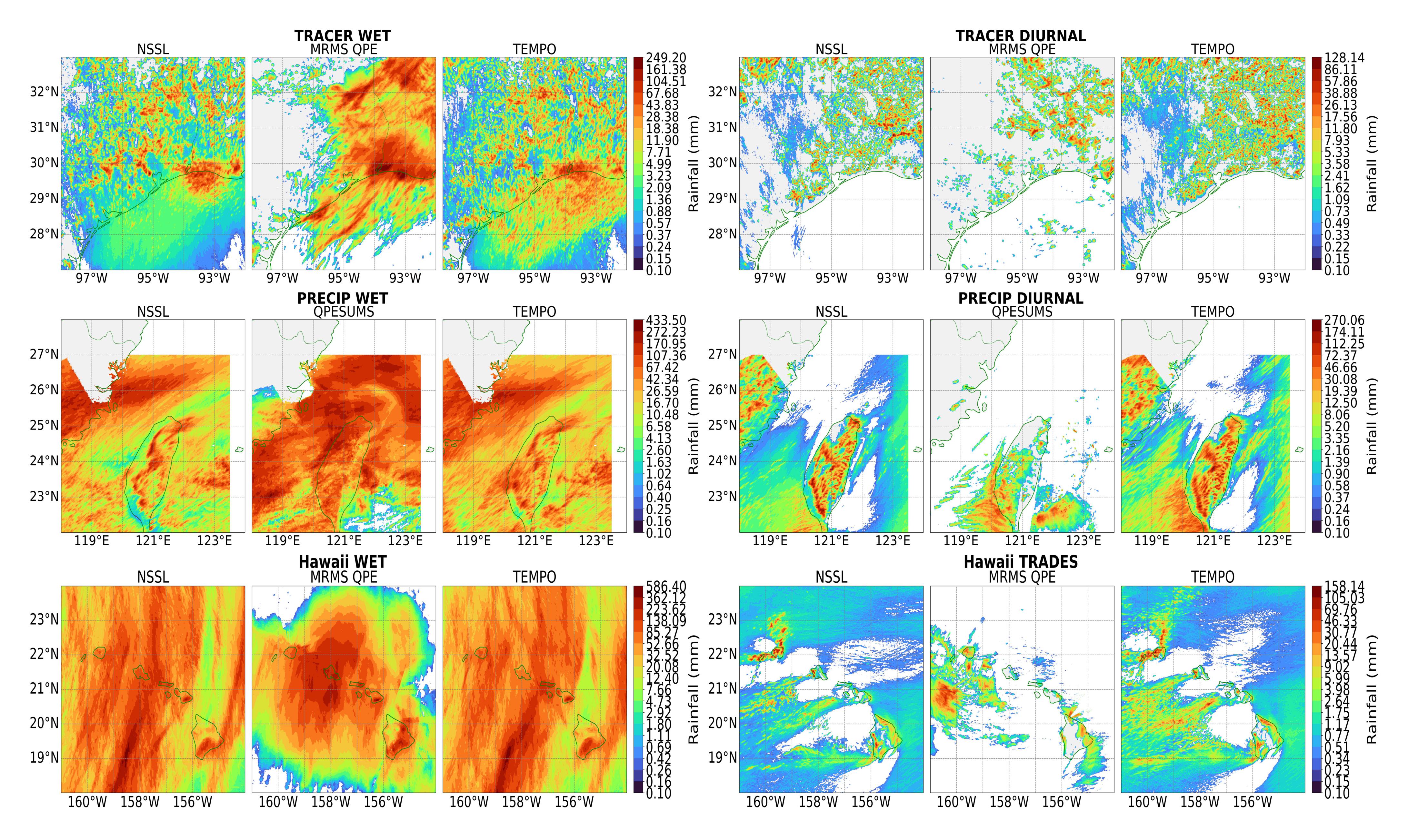}
 \caption{Accumulated rainfall fields from simulation-hour twelve to the end of each simulation. Rows correspond to TRACER (top), PRECIP (middle), and Hawai\okina i (bottom) cases, with WET cases shown in the left column and DIURNAL cases (TRADES for Hawai\okina i) shown in the right column. Within each panel, NSSL (left), observations (center), and TEMPO (right) are shown. To better distinguish meaningful rainfall regions, all data below 0.3 mm were filtered out to remove noise.}\label{figure3}
\end{figure*}

\begin{figure*}[t]
 \noindent\includegraphics[width=\textwidth,angle=0,trim=0 9 0 0]{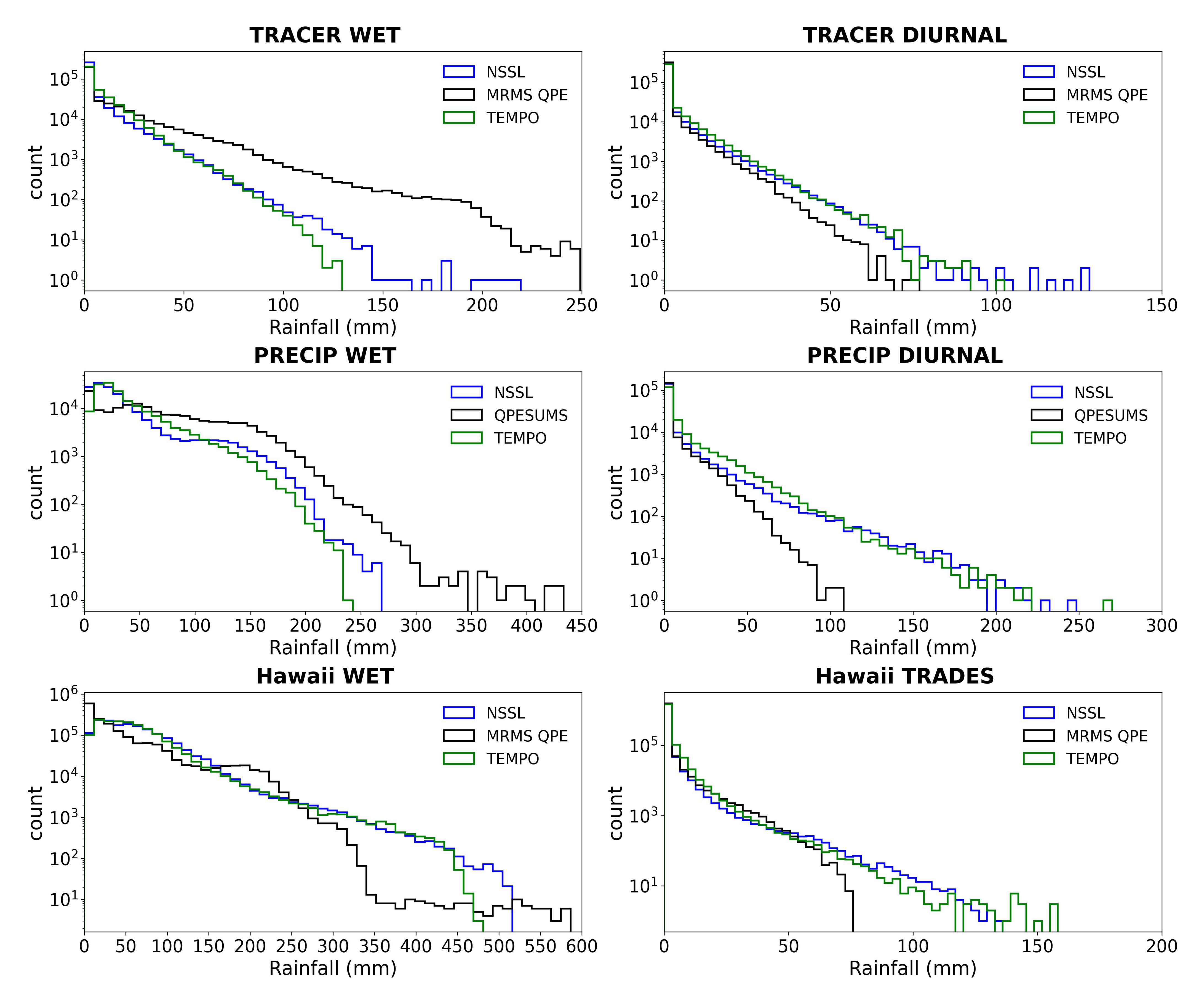}
 \caption{Histograms of QPE rainfall observations (black) and simulated rainfall after interpolation to the observational grid. (blue: NSSL; green: TEMPO). Rows correspond to TRACER (top), PRECIP (middle), and Hawai\okina i (bottom) cases, with WET cases shown in the left column and DIURNAL cases (TRADES for Hawai\okina i) shown in the right column.}\label{figure4}
\end{figure*}

\begin{figure*}[t]
 \noindent\includegraphics[width=\textwidth,angle=0,trim=0 9 0 0]{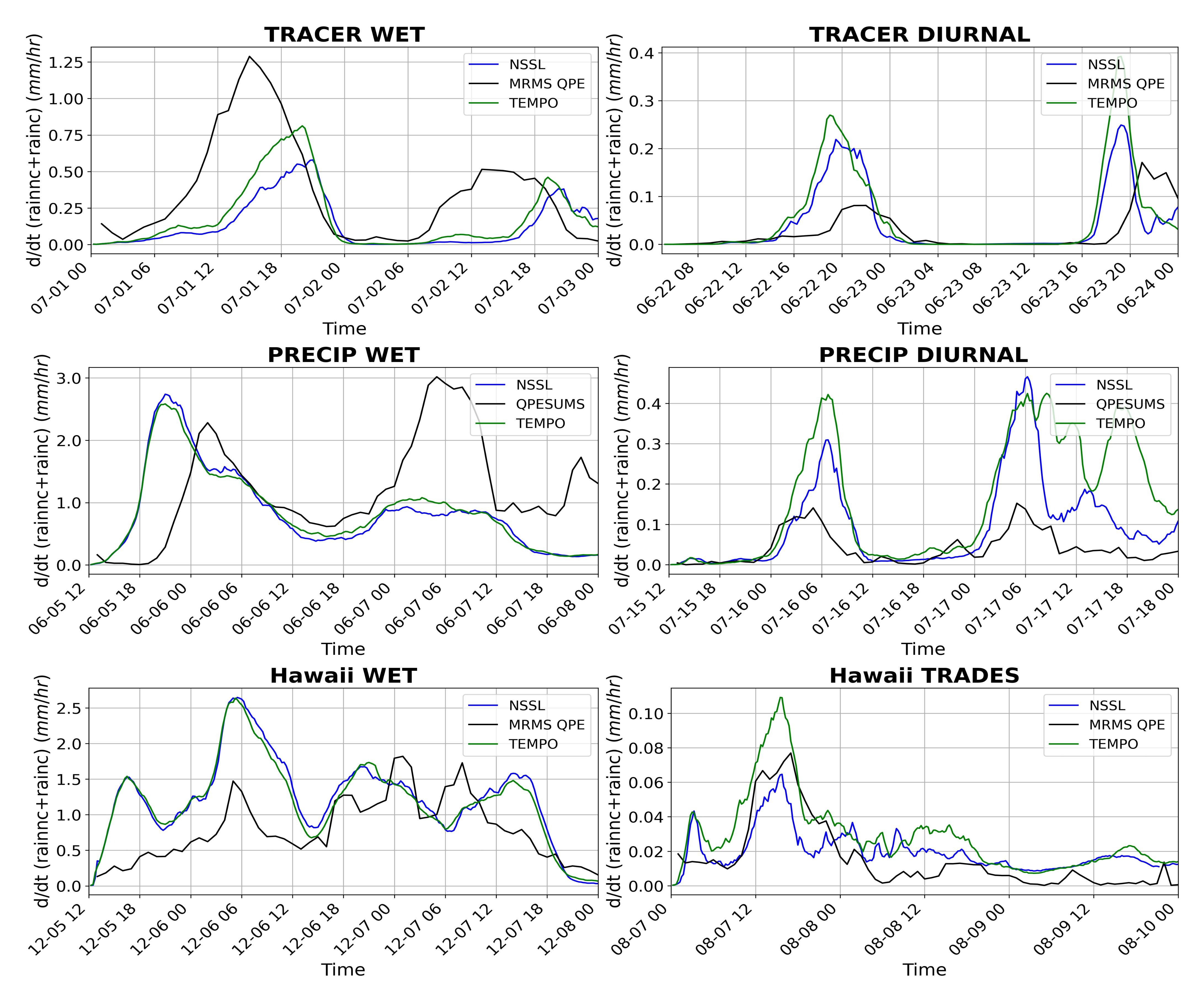}
 \caption{Area-averaged rain rate time series (UTC) from QPE observations (black) and model simulations at the nearest grid point to each station (blue: NSSL; green: TEMPO). Rows correspond to TRACER (top), PRECIP (middle), and Hawai\okina i (bottom) cases, with WET cases shown in the left panels and DIURNAL cases (TRADES for Hawai\okina i) shown in the right panels.}\label{figure5}
\end{figure*}

\begin{figure*}[t]
 \noindent\includegraphics[width=\textwidth,angle=0,trim=0 9 0 0]{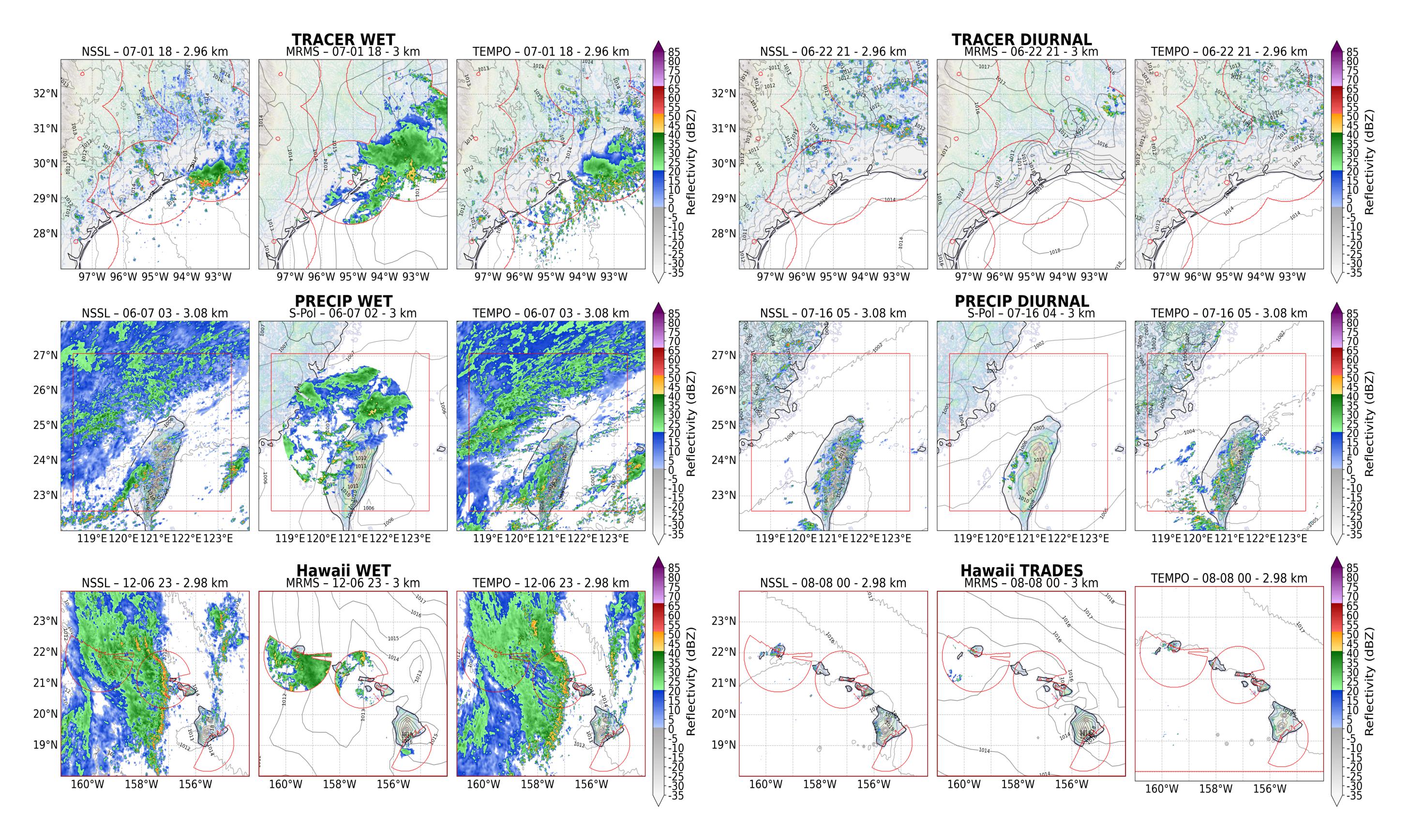}
 \caption{Horizontal fields of radar reflectivity at 3 km altitude at selected times (UTC) representative of dominant convective structures and organizational differences exhibited by each regime. Rows correspond to TRACER (top), PRECIP (middle), and Hawai\okina i (bottom) cases, with WET cases shown in the left column and DIURNAL cases (TRADES for Hawai\okina i) shown in the right column. Within each panel, NSSL (left), observations (center), and TEMPO (right) are shown. Red outlines indicate the spatial extent of available radar observations.}\label{figure6}
\end{figure*}

Accumulated rainfall (Figure~\ref{figure3}), accumulated rainfall distributions (Figure~\ref{figure4}), rain-rate time series (Figure~\ref{figure5}), and 3-km reflectivity (Figure~\ref{figure6}) showed consistent inter-scheme differences, though reduced in weakly-forced cases and large-scale strongly-forced events in PRECIP and Hawai\okina i/WET events. TEMPO produced more widespread light-to-moderate rainfall (Figures~\ref{figure3} and \ref{figure4}) and more numerous, weaker convective cores (Figure~\ref{figure6}), better capturing stratiform convection, consistent with \cite{Feng2018}, most clearly in TRACER/WET, where it improved ocean rainfall estimates that are largely absent in NSSL (Figure~\ref{figure3}). In contrast, NSSL produced more localized, intense rainfall with higher accumulations over smaller areas (Figure~\ref{figure3}), a heavier upper-tail rainfall distribution indicating less frequent, but more intense events (Figure~\ref{figure4}), and stronger, broader (“blobbier”) convective cores (Figure~\ref{figure6}). Specifically, while convective cores were between 40 and 50 dBZ for TEMPO, NSSL’s cores were somewhat over 50 dBZ. Even in regions where NSSL produced scattered convection similar to TEMPO, cores remain slightly broader and stronger than those in TEMPO. NSSL also showed higher reflectivity along line-like convective bands shared by both schemes in Hawai\okina i/WET. The heavy upper-tail was smaller in Hawai\okina i/TRADES and more similar between schemes in weakly-forced cases (Figure~\ref{figure4}). Although NSSL had more intense rainfall, TEMPO generally produced higher average rainfall (e.g. 0.25--0.5$\ \mathrm{mm\,hr^{-1}}$ in TRACER cases) and began raining earlier (e.g. 1--2 hrs in TRACER cases) (Figure~\ref{figure5}). Removing convective-scheme rainfall (not shown) confirmed that rainfall was dominated by microphysics, with convective scheme rainfall contributing only a small fraction of the total as weak, broad background rainfall, mainly in TRACER/WET, PRECIP/DIURNAL, and Hawai\okina i/TRADES.

Although inter-scheme contrasts exist, shared structural biases relative to observations were more pronounced. Both schemes produced overly scattered (“popcorn”-like) precipitation and reflectivity compared to observations (Figures~\ref{figure3} and \ref{figure6}), which instead showed larger, more contiguous (organized) convective regions and broader stratiform areas. This bias was especially pronounced in weakly-forced cases and regions with complex mountainous terrain where simulations produced numerous small, dispersed cells, rather than fewer, more isolated cells. This tendency was more pronounced with TEMPO since its cores already tended to be more scattered than NSSL (Figure~\ref{figure6}). In TRACER/WET, this lack of organization was partly linked to the model’s failure to maintain the coastal low, although the general convective evolution and location were still reasonably captured (Figures~\ref{figure3} and \ref{figure6}). Overall, simulations struggled to represent the full spectrum of convective organization, underrepresenting larger organized systems while overproducing weak, scattered convection. Both schemes also exhibited regime-dependent biases in rainfall frequency and magnitude. Rainfall histograms (Figure~\ref{figure4}) and rain-rate time series (Figure~\ref{figure5}) show that simulations generally underestimated rainfall and peak rates in WET cases, but overestimated them in DIURNAL cases, with Hawai\okina i/WET as an exception.

\subsection{Spatial Metrics and Vertical Distribution}

\begin{figure*}[t]
 \noindent\includegraphics[width=\textwidth,angle=0,trim=0 9 0 0]{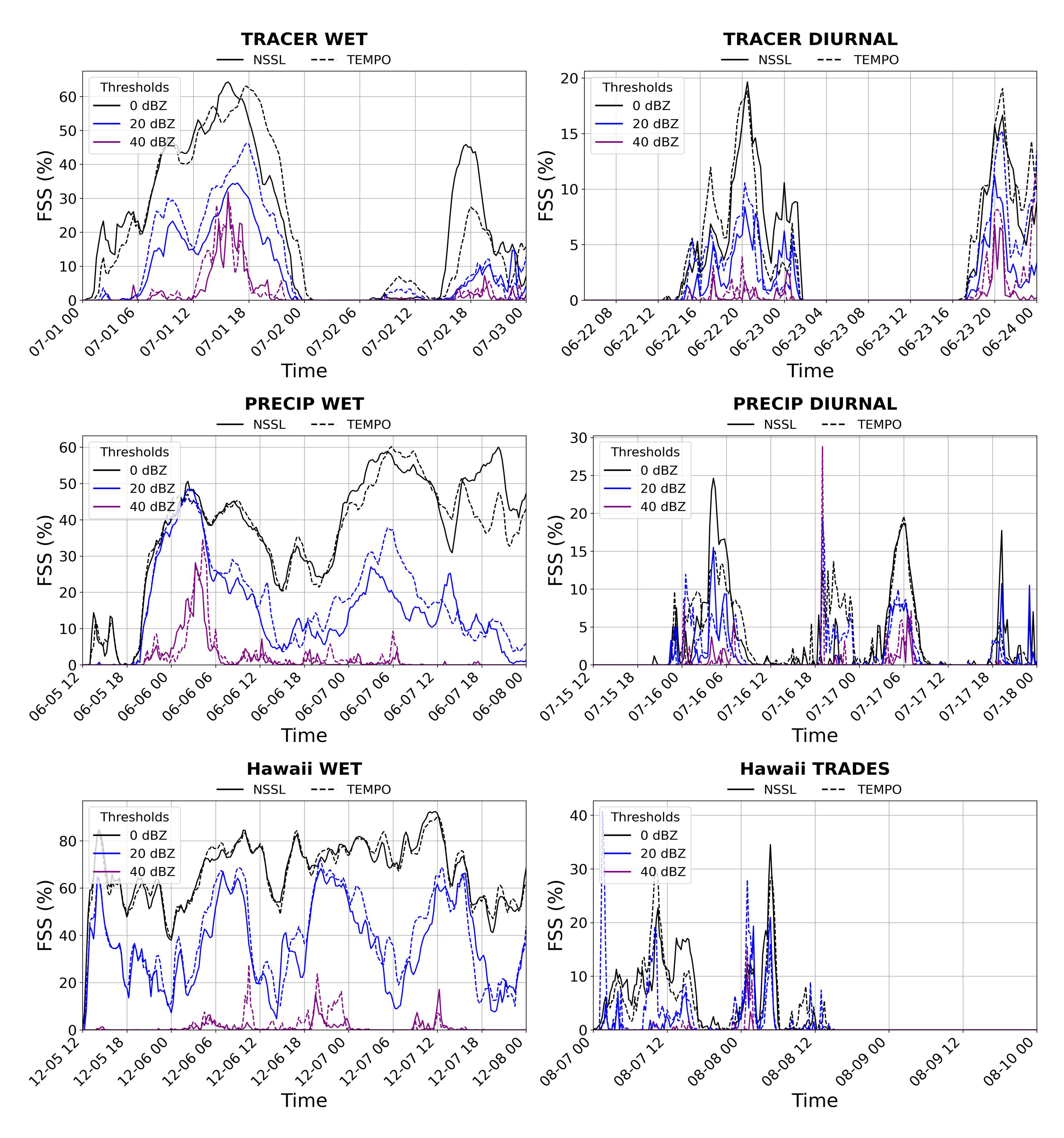}
 \caption{Fractions Skill Score (FSS) using a 5-km window at 3-km altitude evaluated at each time (UTC) from simulated and observed reflectivity across multiple reflectivity thresholds ($> 0$, $\ge 20$, and $\ge 40$) for each case. Scores range from 0 to 100\%, with higher values indicating better agreement. Rows correspond to TRACER (top), PRECIP (middle), and Hawai\okina i (bottom) cases, with WET cases shown in the left column and DIURNAL cases (TRADES for Hawai\okina i) shown in the right column.}\label{figure7}
\end{figure*}

\begin{figure*}[t]
 \noindent\includegraphics[width=\textwidth,angle=0,trim=0 9 0 0]{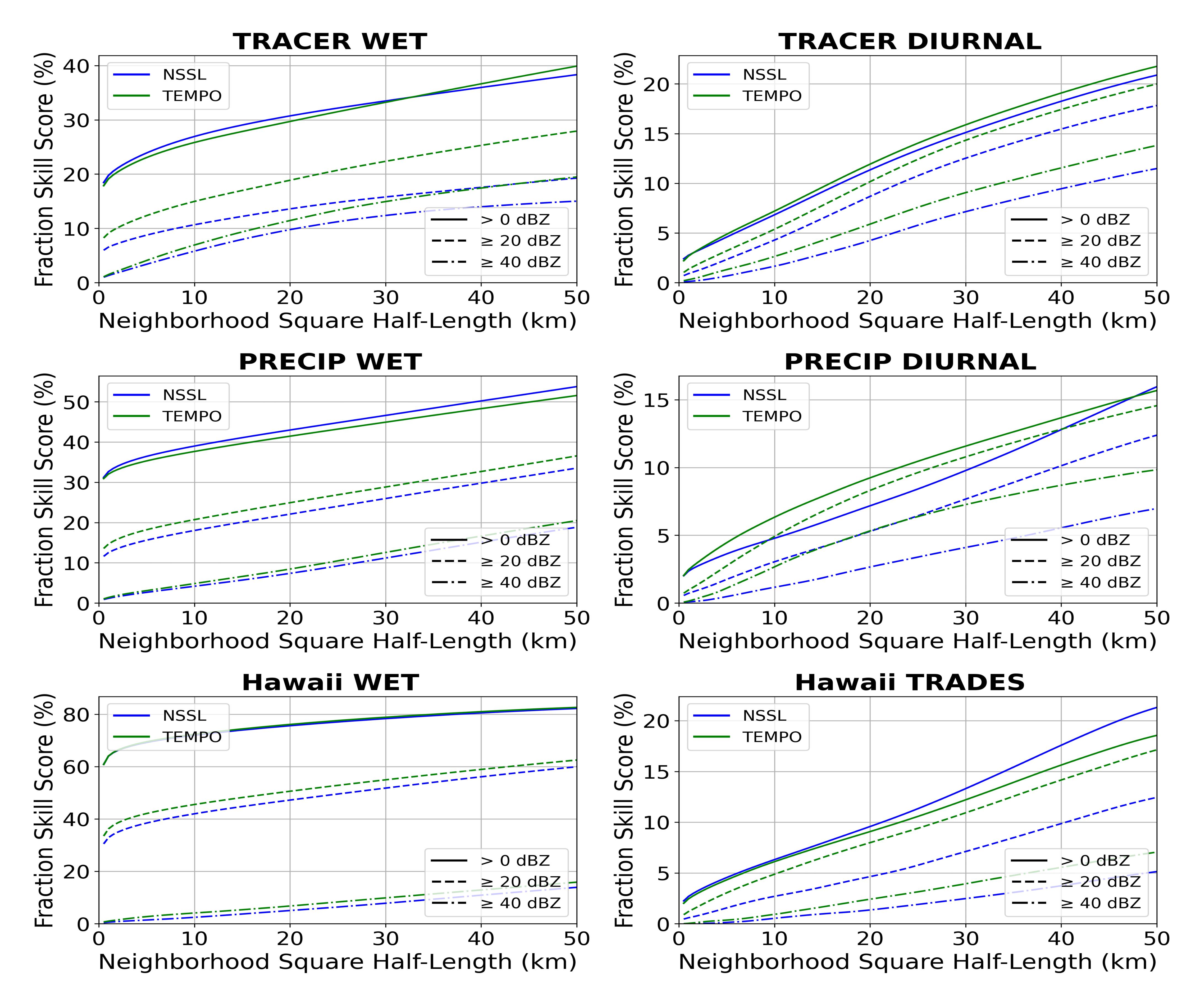}
 \caption{Same as Figure 7, instead averaged over time at 3 km and binned by the neighborhood square half-length for the window employed in the FSS algorithm.}\label{figure8}
\end{figure*}

FSS at 3-km altitude, shown both as a 5-km-window time series (Figure~\ref{figure7}) and as values binned by window size across all times (Figure~\ref{figure8}), provides a quantitative assessment of inter-scheme differences. Inter-scheme differences in FSS were small, depending more on reflectivity threshold and weather regime. TEMPO often had slightly better skill by about 5\% on average for $\ge 20$ dBZ and $\ge 20$ dBZ consistently throughout neighborhood sizes tested up to 50 km, though both schemes did not show a consistent advantage at $>0$ dBZ across cases. Skill was highest at low thresholds ($>0$ dBZ), reflecting convective coverage, and decreased for stronger cores ($\geq 40$ dBZ). Both schemes showed similar regime-dependent behavior with higher performance was observed in strong-forced regimes, especially for the large-scale cases, and lower performance in weakly-forced regimes, with differences ranging from 20-60\%.

\begin{figure*}[t]
 \noindent\includegraphics[width=\textwidth,angle=0,trim=0 9 0 0]{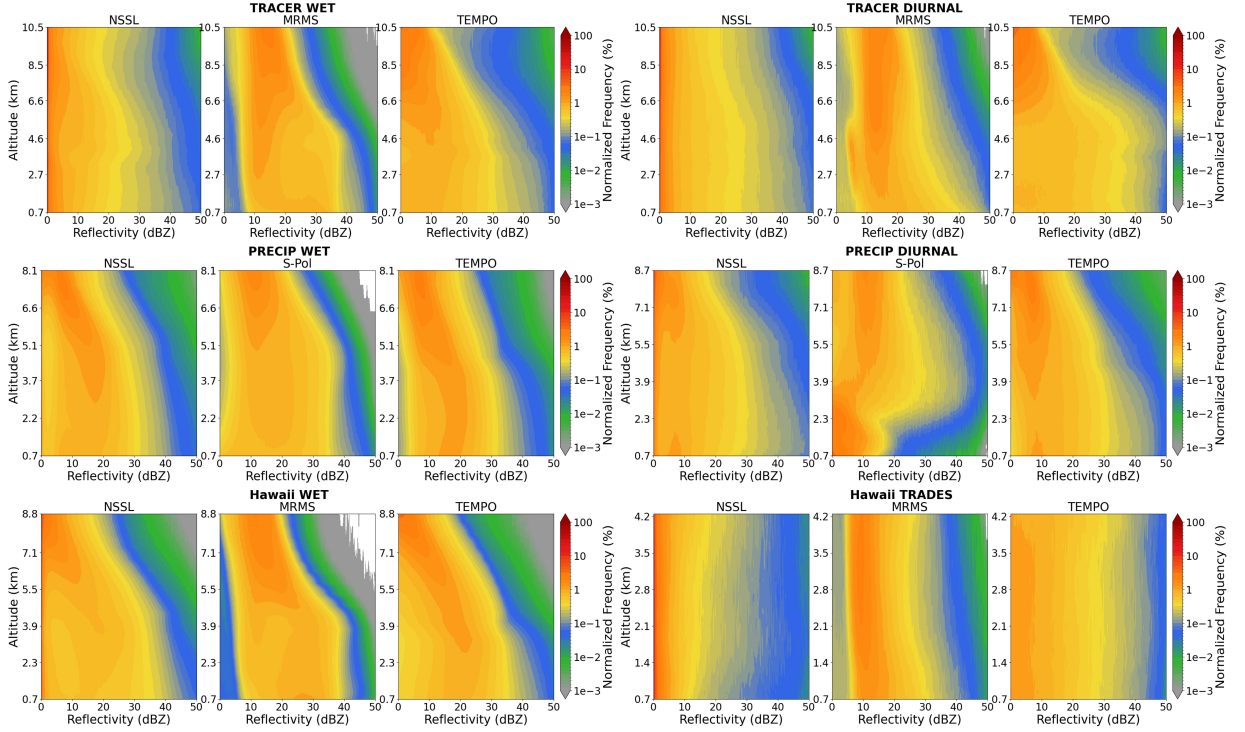}
 \caption{Contoured frequency by altitude diagrams (CFADs) of radar reflectivity binned over all times, normalized at each height, for all cases. Rows correspond to TRACER (top), PRECIP (middle), and Hawai\okina i (bottom) cases, with WET cases shown in the left column and DIURNAL cases (TRADES for Hawai\okina i) shown in the right column. Within each panel, NSSL (left), observations (center), and TEMPO (right) are shown. Observations for the TRACER and Hawai\okina i cases are from MRMS reflectivity mosaic data, while observations for the PRECIP cases are from gridded S-Pol radar data collected during the PRECIP campaign.}\label{figure9}
\end{figure*}

\begin{figure*}[t]
 \noindent\includegraphics[width=\textwidth,angle=0,trim=0 9 0 0]{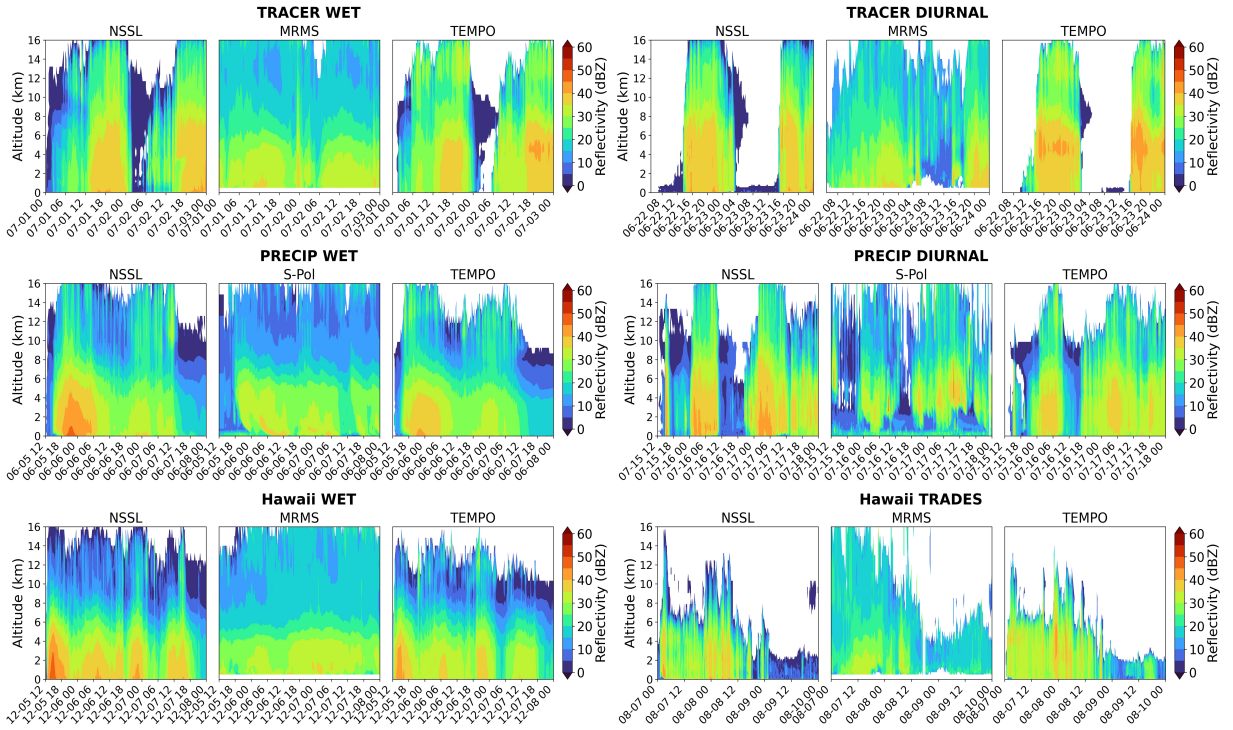}
 \caption{Area-averages of radar reflectivity for model and observations for each time (UTC). Rows correspond to TRACER (top), PRECIP (middle), and Hawai\okina i (bottom) cases, with WET cases shown in the left column and DIURNAL cases (TRADES for Hawai\okina i) shown in the right column. Within each panel, NSSL (left), observations (center), and TEMPO (right) are shown. Low-level reflectivity area-averages were calculated by averaging reflectivity from 0--3 km MSL, consistent with the 3-km FSS level. Reflectivity averages were computed in linear Z-space before conversion to dBZ.}\label{figure10}
\end{figure*}

CFADs (Figure~\ref{figure9}) and area-averaged radar reflectivity profiles (Figure~\ref{figure10}) were used to examine vertical convective structure. Inter-scheme differences are again minor compared to shared biases relative to observations. Both schemes broadly captured the vertical reflectivity structure, but weak values (0--10 dBZ) were overrepresented at all levels, shifting the distribution toward weaker reflectivity most clearly above 3 km. NSSL also struggled to reproduce observed reflectivity above the melting level, consistent with previous studies \citep[e.g.,][]{Varble2011,Song2022}. NSSL also produced broader reflectivity distributions and weaker vertical gradients than TEMPO and observations, consistent with stronger, deeper convective cores. Hawai\okina i/TRADES was an exception, where both schemes and observations show weak, vertically-uniform reflectivity. Compared to observations, both schemes produced storms that were too vertically deep (more so with NSSL) and failed to capture weakening aloft. Simulations also varied much more over time than in the relatively steady observations. Both simulations also showed a stronger diurnal signal than observed, often suppressing convection overnight when observations indicated continued reflectivity. These clear inter-scheme differences motivate further analysis of the dynamical and microphysical fields.

\subsection{Area-Average Comparisons}

\begin{figure*}[t] \centering \includegraphics[ width=0.95\textwidth, height=0.85\textheight, keepaspectratio, trim=0 9 0 0, clip ]{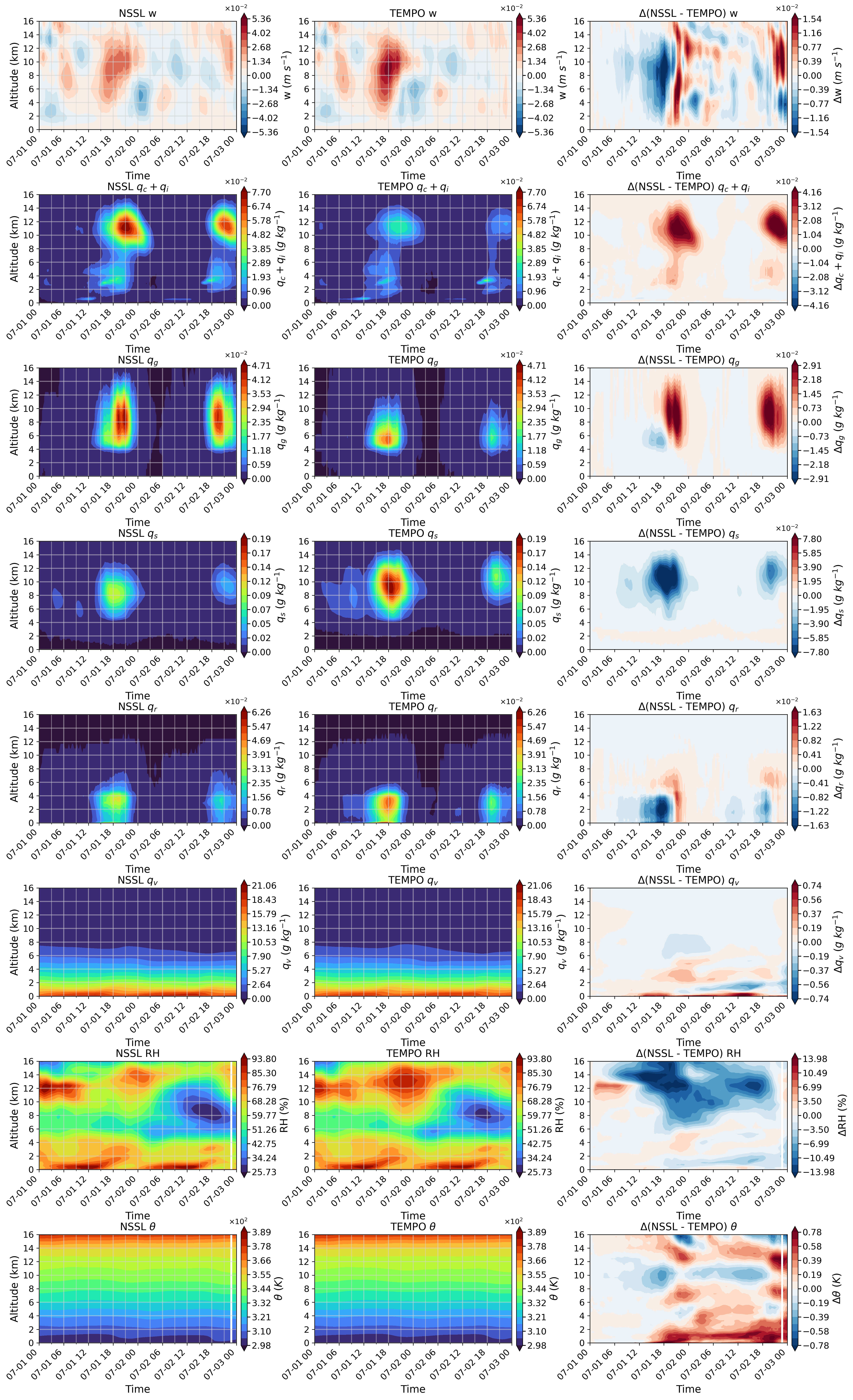}
 \caption{Shows area-averages of selected model variables for each time (UTC) for the TRACER/WET case. Each row is for a different variable, including vertical velocity ($w$), cloud water and ice mixing ratios ($q_c+q_i$), graupel mixing ratio ($q_g$), snow mixing ratio ($q_s$), rain mixing ratio ($q_r$), water vapor ($q_v$), relative humidity ($RH$), and potential temperature ($\theta$). Figures for NSSL (left), TEMPO (center), and NSSL-TEMPO (right) are shown. All plots use color limits based on the 99th percentile of the plotted variable.}\label{figure11}
\end{figure*}

\begin{figure*}[t] \centering \includegraphics[ width=0.95\textwidth, height=0.85\textheight, keepaspectratio, trim=0 9 0 0, clip ]{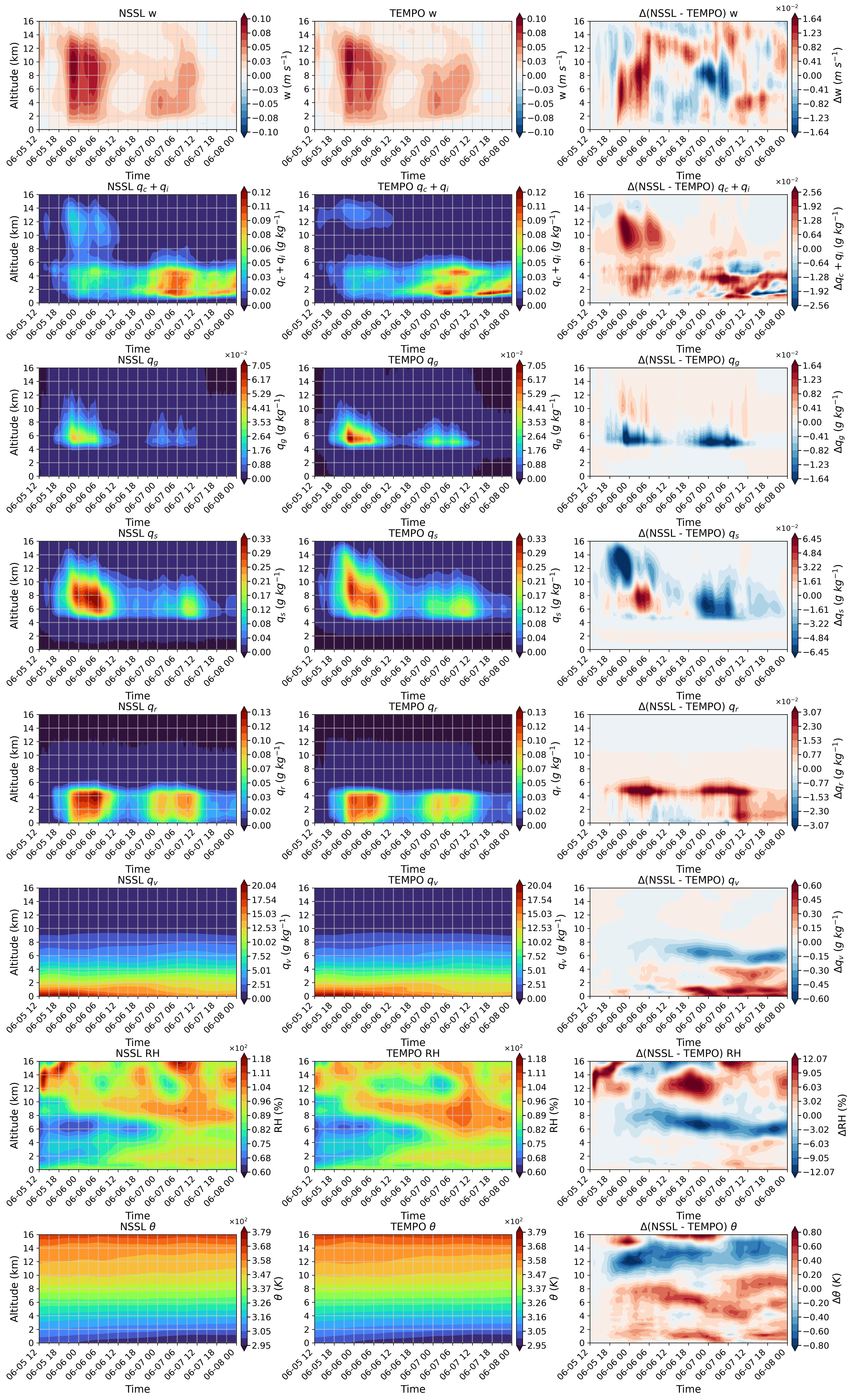}
 \caption{Same as Figure 11, but for the PRECIP/WET case.}\label{figure12}
\end{figure*}

\begin{figure*}[t] \centering \includegraphics[ width=0.95\textwidth, height=0.85\textheight, keepaspectratio, trim=0 9 0 0, clip ]{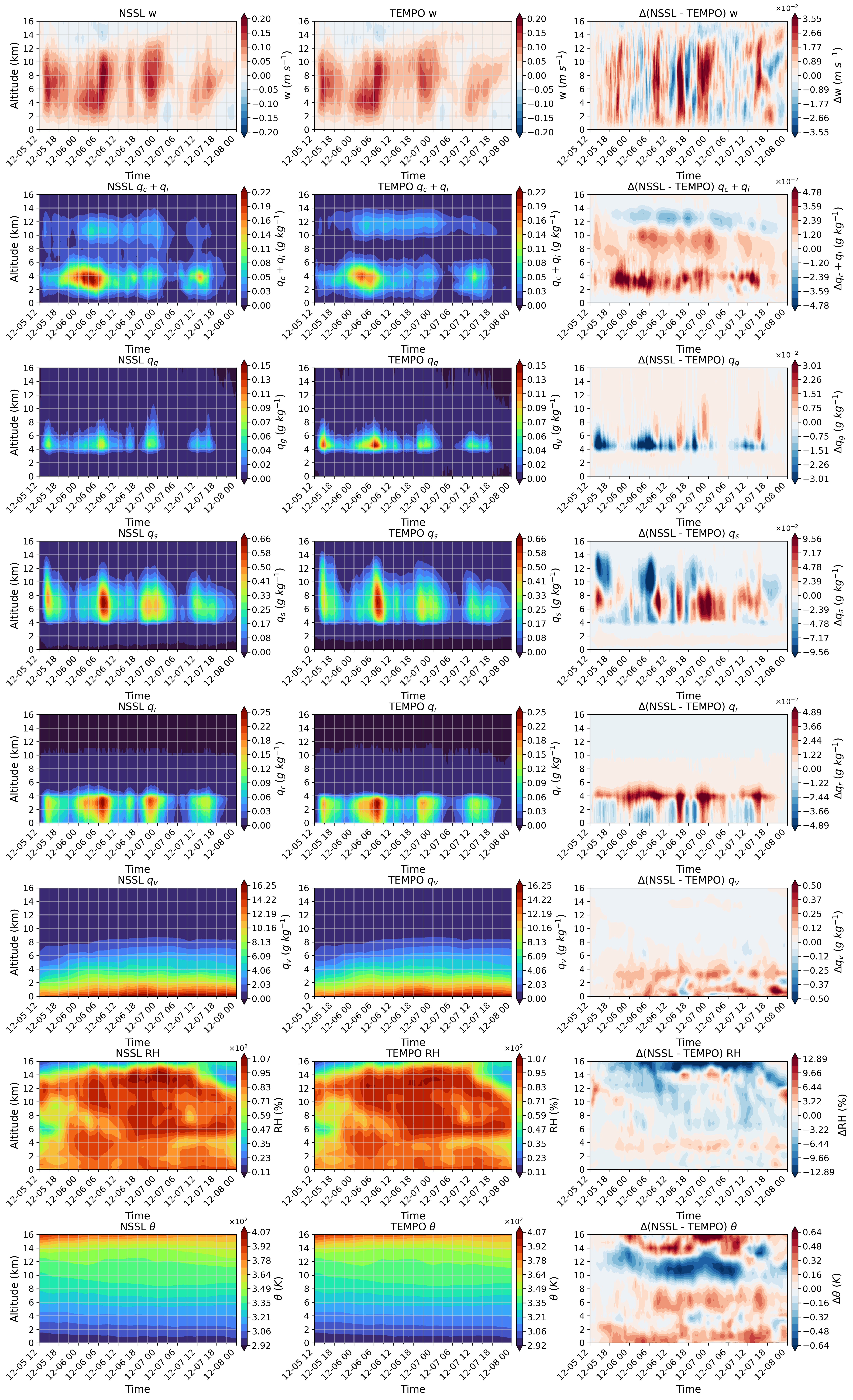}
 \caption{Same as Figure 11, but for the Hawai\okina i/WET case.}\label{figure13}
\end{figure*}

To further investigate reflectivity differences, horizontally-averaged convective diagnostics were examined. Figures~\ref{figure11}--\ref{figure13} show area-averaged time-height diagrams of selected variables for TRACER, PRECIP, and Hawai\okina i cases. Vertical velocity differences showed no clear trend, with either scheme stronger at different times. More robust differences appeared in microphysical structure: NSSL produced deeper convection, with enhanced upper-level condensate/ice (e.g., 99th percentile $4.16 \times 10^{-2}\ \mathrm{g\,kg^{-1}}$ greater in TRACER/WET) and graupel (e.g., $2.91 \times 10^{-2}$), while TEMPO produced greater snow mixing ratios (e.g., $7.8 \times 10^{-2}$) extending higher, often at the expense of graupel. This is consistent with previous studies comparing Thompson-type schemes with other two-moment schemes \citep{Weverberg2013,Guo2019,Song2022,Johnson2023}. Some exceptions occured: in PRECIP/WET, NSSL briefly had more snow before TEMPO had more later, while in Hawai\okina i/WET, NSSL had more mid-level snow, but TEMPO maintained greater upper-level snow (10--14 km). Rain mixing ratios also differed. NSSL produced deeper rain regions with greater mixing ratios (e.g., $\sim\!1.02 \times 10^{-2}\ \mathrm{g\,kg^{-1}}$) aloft ($\sim$4--6 km), while TEMPO had longer-lasting rain cores with greater mixing ratios (e.g., $1.63 \times 10^{-2}\ \mathrm{g\,kg^{-1}}$) at lower levels (0--4 km). TEMPO had cooler (e.g., $0.78$ K) temperatures below 10 km (2 km in Hawai\okina i/TRADES due to strong stability), reflecting enhanced evaporative cooling associated with more widespread rainfall. There weren't robust relative humidity differences between schemes, though NSSL was sometimes slightly higher at low levels by a few percentage points, also depending on the case. NSSL’s mid- to upper-tropospheric warming (5--10 km) and drier conditions above $\sim$6 km are consistent with enhanced latent heating from ice-phase deposition. NSSL’s cooler temperatures above ~10 km may be related to scheme-dependent differences in upper-level cloud and anvil structure. Cold-pool characteristics, identified using density potential temperature perturbations $< -1$ K and near-surface rainwater mixing ratios $1 \times 10^{-6}$, lacked consistent scheme-dependent differences, possibly because differences in storm evolution obscured systematic microphysical effects, although NSSL occasionally produced stronger negative $\theta_v$ perturbations suggesting locally stronger cold-pools.

\begin{figure*}[t]
 \noindent\includegraphics[width=\textwidth,angle=0,trim=0 9 0 0]{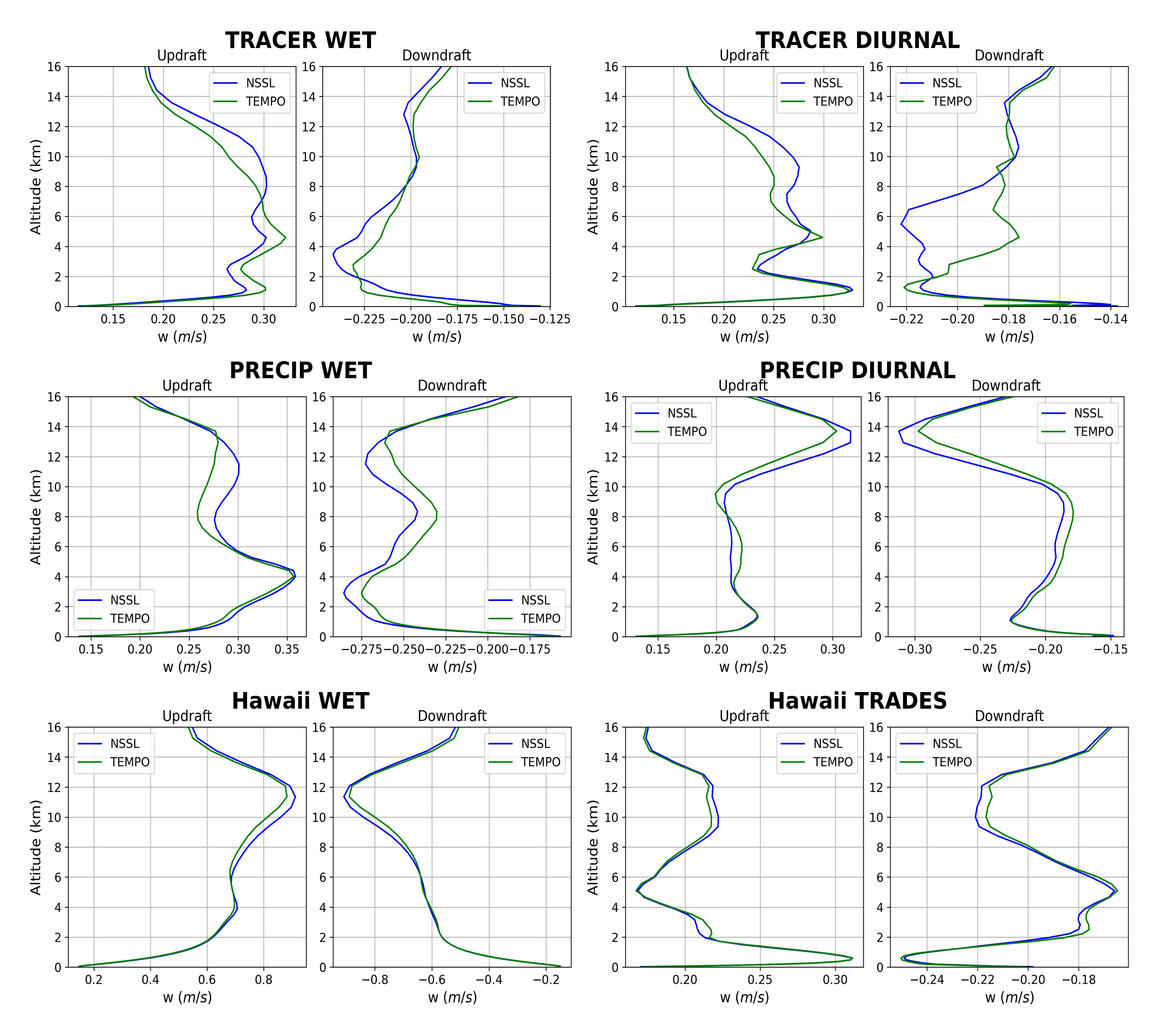}
 \caption{Thresholded area-averaged vertical profiles of vertical velocity for boundary layer and aloft updrafts ($w > 0.1\ \mathrm{m\,s^{-1}}$) and downdrafts (($w < -0.1\ \mathrm{m\,s^{-1}}$) averaged over all times. Rows correspond to TRACER (top), PRECIP (middle), and Hawai\okina i (bottom) cases, with WET cases shown in the left column and DIURNAL cases (TRADES for Hawai\okina i) shown in the right column. Within each panel, updraft (left) and downdraft (right) profiles are shown.}\label{figure14}
\end{figure*}

\begin{figure*}[t]
 \noindent\includegraphics[width=\textwidth,angle=0,trim=0 9 0 0]{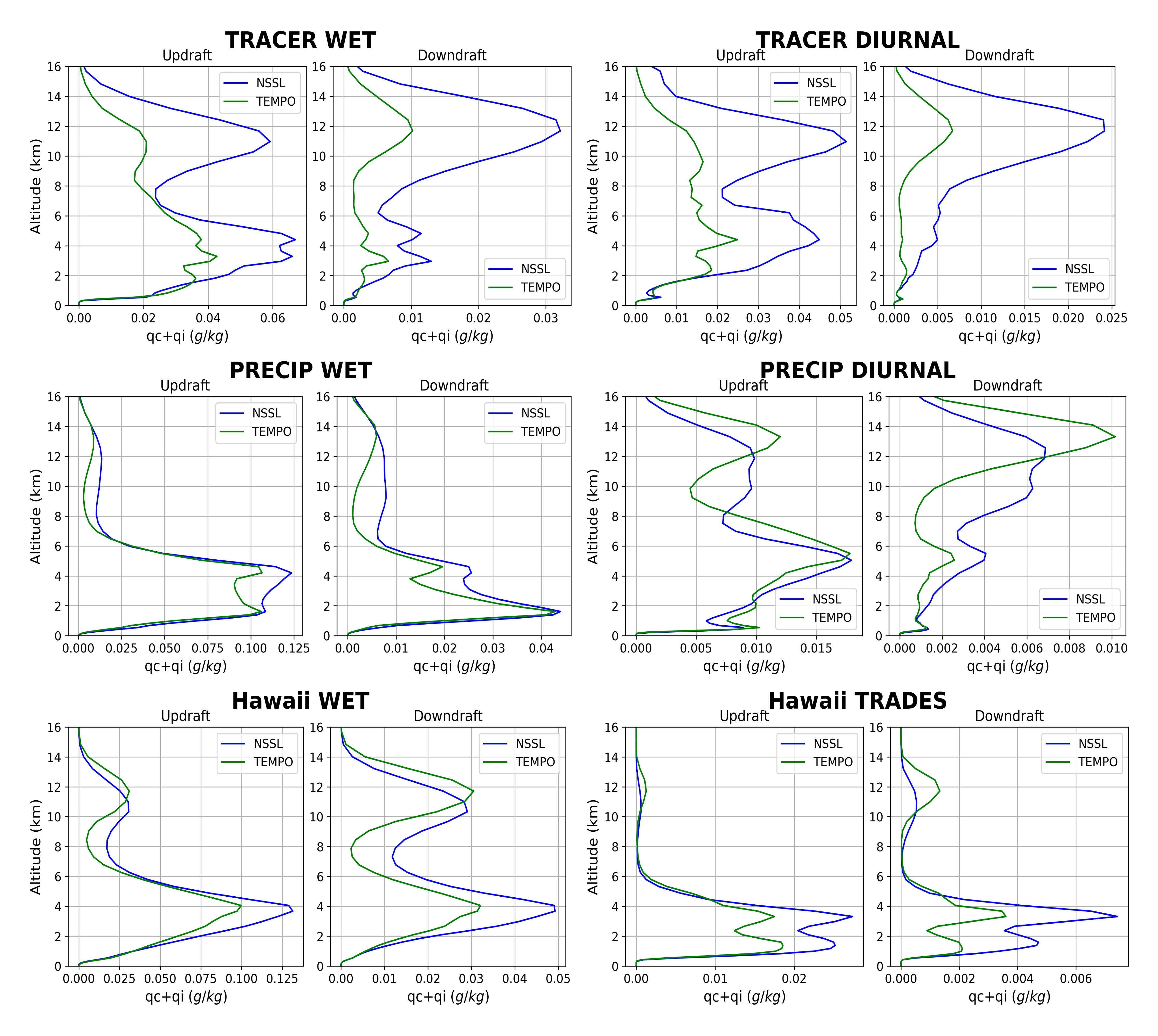}
 \caption{Same as Figure 14, but for cloudiness ($q_c+q_i$).}\label{figure15}
\end{figure*}

To examine inter-scheme differences in convective core updraft structure, thresholded area-averaged profiles were computed using standard vertical velocity criteria (Figures ~\ref{figure14}--\ref{figure15}). A lower threshold of $0.1\ \mathrm{m\,s^{-1}}$ was chosen over higher values (e.g., $0.5\ \mathrm{m\,s^{-1}}$) and condensate was not included to retain a broader range of vertical motions, including boundary-layer updrafts, cloudy updrafts, and turbulent updrafts aloft. Additional hydrometeor-content thresholds (not shown) produced similar downdraft, but inconsistent updraft results, because scheme-dependent storm structures caused microphysical thresholds to sample different storm regions and reduce comparability. Primarily in WET cases, TEMPO produced comparable or slightly stronger low-level updrafts and downdrafts, whereas NSSL developed stronger, deeper velocity maxima aloft with enhanced hydrometeor content. Velocity differences between NSSL and TEMPO were modest, about $0.0025$--$0.05\ \mathrm{m\,s^{-1}}$, whereas hydrometeor-content differences were more clear, about $0.02$--$0.04\ \mathrm{g\,kg^{-1}}$, consistent with increased latent heating from ice-phase processes with NSSL.

\begin{figure*}[t]
 \noindent\includegraphics[width=\textwidth,angle=0,trim=0 9 0 0]{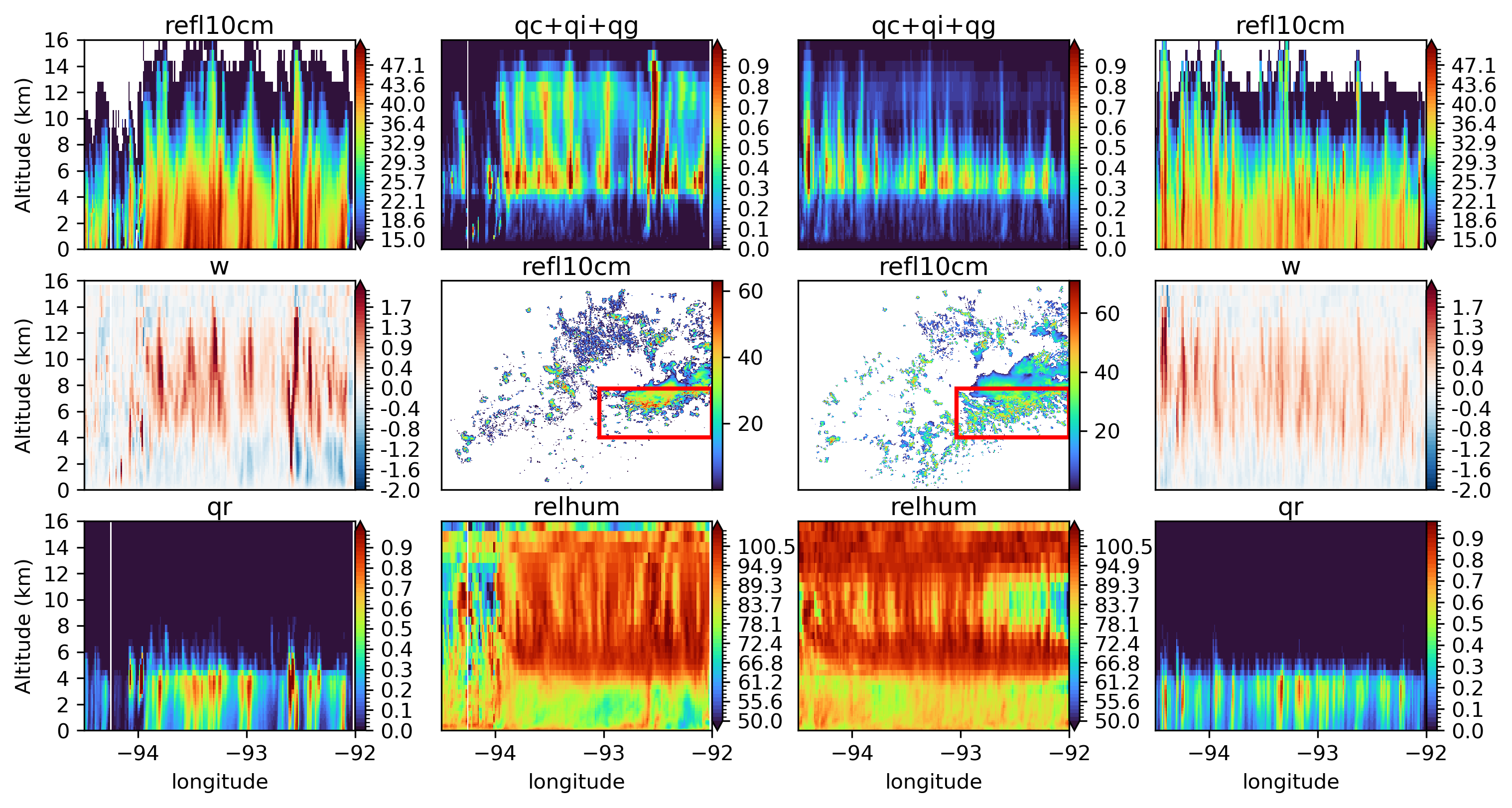}
 \caption{Horizontal (3 km altitude) and vertical cross sections of selected variables from zoomed-in regions of representative convective systems within individual snapshots shown in Figure 6, illustrating differences in convective core structure between the NSSL (left) and TEMPO (right) simulations. Shown for TRACER/WET case. All variables are in standard units, mixing ratios are in units of $\mathrm{g\,kg^{-1}}$. All plots use color limits based on the 99th percentile of the plotted variable.}\label{figure16}
\end{figure*}

\begin{figure*}[t]
 \noindent\includegraphics[width=\textwidth,angle=0,trim=0 9 0 0]{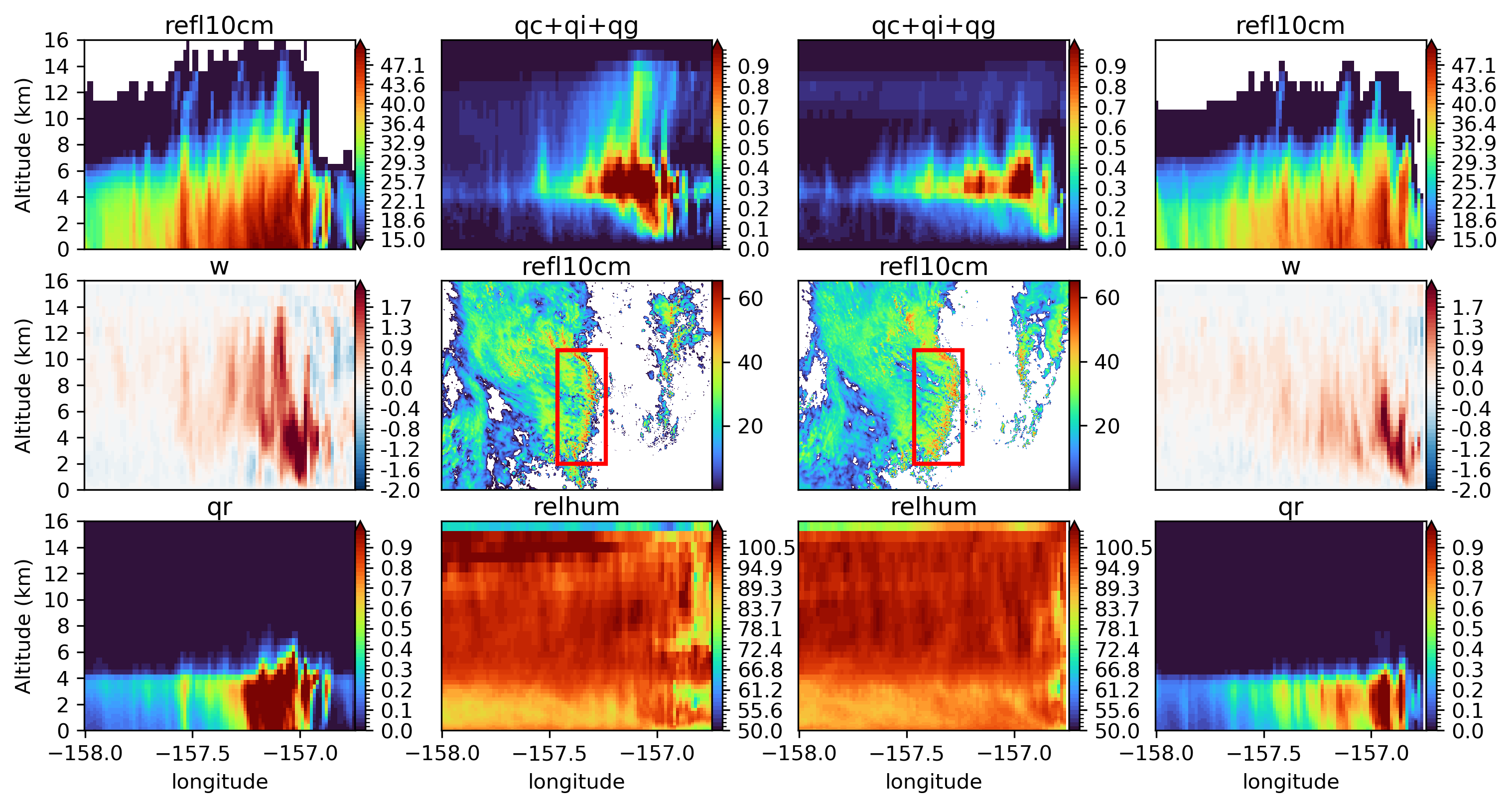}
 \caption{Same as Figure 16, but for the Hawai\okina i/WET.}\label{figure17}
\end{figure*}

To further investigate structural differences at a local storm scale, representative storms from Figure~\ref{figure6} were analyzed using 3-km horizontal and vertical cross sections. A reflectivity threshold of $Z > 15$ dBZ at 3 km was applied to isolate convective and stratiform regions while filtering weak convection, following \cite{Putnam2017}. Examples include a coastal storm in TRACER/WET (Figure~\ref{figure16}) and a squall line in Hawai\okina i/WET (Figure~\ref{figure17}). NSSL again exhibited broader, stronger updrafts (e.g. $\sim\!1\,\mathrm{m\,s^{-1}}$ in TRACER/WET), larger reflectivity (e.g. $\sim\!10\,\mathrm{dBZ}$), and greater cloud and ice aloft (e.g. $\sim\!1\,\mathrm{g\,kg^{-1}}$), while TEMPO produced weaker, but more scattered convection. At the local storm scale, NSSL produced higher rain mixing ratios (e.g. $0.5\,\mathrm{g\,kg^{-1}}$) in the lowest $\sim$4 km, while TEMPO was more widespread horizontally. This contrasts with the domain-averaged rainfall, where TEMPO had higher values. Together, these results suggest that NSSL produces more intense local precipitation, while TEMPO produces broader, lower-level rainfall coverage on average.

\section{Summary and Discussion}

This study evaluated how two operational bulk microphysics schemes, NSSL and TEMPO, influence convection and precipitation in high-resolution, convection-permitting Regional-MPAS simulations. Experiments spanned multiple subtropical/tropical regions in Houston, Taiwan, and Hawai\okina i under strongly-forced and weakly-forced regimes. MPAS-A reproduced the large-scale environment and diurnal evolution well, though notable storm-scale differences appeared between schemes and more so relative to observations.

Both schemes struggled to reproduce convective organization and intensity seen in observations. Simulations exhibited a tendency towards numerous, scattered convective cells and insufficient stratiform regions. This limitation in the convective-stratiform partition is well documented in kilometer-scale cloud-resolving simulations \citep[e.g.,][]{Zhou2007,Luo2010,Varble2011}. \cite{Zhou2007} attributed this less efficient conversion into stratiform rain to excessive snow and graupel aloft with slower fallout and melting. It was found that NSSL favors fewer stronger and broader cores and convective updrafts, with higher ice and graupel aloft relative to dry air. These higher amounts of ice hydrometeors may involve slower ice-phase growth and delay fallout may retain more condensate aloft \citep[e.g.,][]{Weverberg2013} and reduce precipitation efficiency \citep[as defined in][]{Sui2007}. This helps explain NSSL’s lower domain-averaged rain mixing ratios despite stronger localized rainfall maxima: NSSL has more ice and graupel aloft within fewer, deeper cores, while TEMPO’s reduced ice/graupel support broader stratiform precipitation, and more efficient low-level rainfall. TEMPO also had oceanic rainfall in TRACER cases, which was largely absent with NSSL, further supporting this idea. \cite{Song2018} also found larger mean raindrop sizes in Thompson compared to NSSL, which can increase rain mixing ratios. Other possible contributing factors include uncertainties in ICs, insufficient resolution, and limitations in representing horizontal turbulent mixing and cold-pool dynamics. For example, limited lateral mixing aloft may suppress merging of convective elements \citep[e.g.,][]{Zhang2021} or moist detrainment leaving convection overly isolated and cellular. Similar to the former, \cite{Bengtsson2012} showed that stronger horizontal diffusion reduces small convective cores, suggesting that weak grid-cell communication leads to overly cellular, poorly organized convection. \cite{weisman1997} also found that, in kilometer-resolution simulations, unresolved convection can form weaker cold-pools, delay initiation, and affect early growth into larger, more organized systems. Further supporting the role of cold-pool dynamics, \cite{Freitas2024} showed that adding a cold-pool gust-front convective parameterization, which uses downdraft MSE deficits to trigger new convection along cold-pool edges, improved mesoscale convective system organization. Future forecasting systems may benefit from moving toward LES-scale resolutions that explicitly resolve turbulent mixing.

In this current study, NSSL substantially underestimated the observed gradient above melting level and exhibited broader reflectivity distributions aloft. This bias is common in microphysics schemes \citep{Varble2011} and specifically in both Thompson and NSSL \citep{Song2022}, though TEMPO did not clearly exhibit this bias here. Sohn and Lim (2022) found larger graupel mixing ratios and smaller graupel number concentrations responsible for overestimated upper-level reflectivity for the NSSL scheme. \cite{Song2018} similarly attributed this to high graupel content across multiple schemes, though found the Thompson scheme produced more realistic upper-level reflectivity owing to its inverse snow density-size relationship, which yields higher concentrations of smaller snow particles. This overestimation is a well-known bias \citep[e.g.,][]{Lang2003,McFarquhar2006}, suggesting process-level refinements to riming/accretion efficiencies, graupel conversion, sublimation, ice nucleation, etc. TEMPO generally outperformed NSSL including more accurately reproducing the observed convective-stratiform partition. This supports that added complexity of two-moment microphysics does not necessarily improve realism \cite[e.g.,][]{Weverberg2013}, thus warranting further evaluation of convective structure in 2-moment schemes.

NSSL’s enhanced cold-rain processes and lower- to mid-tropospheric reflectivity overestimation (0–8 km) (Figure~\ref{figure10}) in both schemes in cases outside the continental US, suggest regional-dependent biases. This is consistent with studies that show reduced performance of schemes tuned for U.S. continent convection when applied to regions that exhibit dominant warm-rain processes \citep[e.g.,][]{Sohn2013,Song2022}. Regional differences may also reflectivity sensitivity to local conditions and drop size distributions \citep{Lamb2026}. Model biases also showed regime dependence. Model rainfall amounts were underestimated (overestimated) in strongly (weakly)-forced cases, with substantially lower spatial agreement in the latter. The underestimation in strongly-forced cases may reflect a bias toward frequent, light precipitation, often linked to excessive drizzle from unresolved subgrid variability \citep{Stephens2010}, as well as reduced precipitation in higher-resolution simulations due to containing more, smaller convective cells that entrain more mid-level air [Bryan and Morrison 2012]. The overestimation in weakly-forced cases may reflect an exaggerated diurnal response, where overly strong surface heating and boundary-layer fluxes promote rapid convective initiation and excessive precipitation \citep{Bechtold2004,Guichard2004,Kirshbaum2011}. Inter-scheme differences were reduced and model skill much improved in strongly forced regimes, such as PRECIP/WET and Hawai\okina i/WET, suggesting greater model skill for large-scale events than locally-driven convective events.

Simulations also had limitations in representing localized weather processes. In TRACER/WET, the model failed to fully capture the evolution of a mesoscale coastal low. In Hawai\okina i/WET, the model produced elongated rainbands instead of the localized rainfall observed over O\okina ahu. Hawaiian rainfall is highly localized and strongly influenced terrain, microclimates \citep{Giambelluca1986}, trade winds \citep{Schroeder1977}, and diurnal circulations \citep{Feng1998,Wang1998}. This suggests model deficiencies in low-level convergence, particularly related to local topographic circulations and flow convergence critical for convective initiation \citep[e.g.,][]{Kirshbaum2014,Wang2015}. Local diabatic heating and the evolving ocean conditions not modeled may have also affected the thermodynamic structure of the Kona low. These limitations in capturing localized convection were even more pronounced in weakly forced-environments. Improving representation of these PBL processes is critical for forecasting convection over islands and complex terrain.

A general limitation of this study is the limited sample size. Although six cases across multiple regions and regimes were analyzed, more cases are needed to robustly assess regime-dependent microphysics performance. The analysis also relied primarily on bulk and structural diagnostics, limiting process-level attribution of the errors. The nonlinear, path-dependent evolution of convection further complicates direct scheme comparisons, as early differences modify the thermodynamic state and influence later development, suggesting targeted restart analyses are needed to isolate inter-scheme differences. Consistent with \cite{Weisman2008}, increased resolution and improved physics can enhance the realism of convective structure and mode, but do not necessarily improve forecast skill in convective initiation timing or location, which in turn affect convective organization. Additionally, limitations in IC resolution and spin-up sensitivity suggest that improved ICs, such as from HRRR forecasts, and data assimilation would greatly enhance forecast accuracy and enable more robust convective comparisons.

\section{Conclusion}

These findings provide guidance for microphysics selection in emerging Regional-MPAS forecasting systems and support inclusion of NSSL and TEMPO microphysics in operational MPAS frameworks. More broadly, this study addresses the challenge of representing convection across diverse subtropical and tropical regions, where warm-rain processes are dominant yet remain poorly captured by current microphysical schemes. Despite reasonable large-scale performance, microphysics choice strongly influences convective structure. NSSL favors localized, intense storms, while TEMPO produces broader convective coverage and higher domain-mean rainfall. Thus, scheme selection should consider both domain-mean effects and localized storm impacts. Beyond these inter-scheme differences, both schemes exhibit organization biases relative to observations that exceed their inter-scheme differences. Simulations often fail to reproduce the observed coherent mesoscale organization, instead producing many scattered cells. The simulations also exhibited regime-dependent errors, failing to reproduce the entire convective spectrum. Rainfall was under- (over-) represented in strongly-forced (weakly-forced) regimes and forecast skill was much lower for weakly-forced regimes; the former may reflect excessive drizzle production or overly abundant small cellular convection, while the latter may reflect an exaggerated diurnal response. Improving the representation of localized precipitation processes is necessary to better capture convection across a wider range of scales. Overall, TEMPO better represented convective organization and vertical structure relative to observations, suggesting that the added complexity of two-moment microphysics does not necessarily improve realism and that convective structure in both schemes should be examined further.

Future efforts should pursue targeted refinements to microphysical parameterizations across various regimes and regions with diverse precipitation characteristics to improve the robust representation of convection and precipitation. Furthermore, iterative analysis and targeted improvements addressing process-level convective biases along the causal evolution of convective systems would benefit prediction of convective organization. Possible avenues include reducing uncertainties in ICs, improving model resolution, and better representation of horizontal mixing, localized moisture convergence, and cold-pools. Improved data assimilation, such as radar assimilation, together with denser observation networks, would strengthen ICs and further improve forecast skill and model verification. Even small errors in convective evolution can rapidly amplify and alter subsequent storm development, making accurate process-level representation essential for reliable high-resolution forecasting and actionable local-scale prediction.

\clearpage
\acknowledgments

The authors would like to thank Ted Mansell and Anders Jensen (NOAA) for providing the model used in this study and for helpful discussions and guidance. We are grateful to Bill Skamarock, Michael Duda, and Ming Chen (NCAR) for their assistance in overcoming compilation and numerical issues encountered during this work. Both Ming Chen and Abishek Gopal also provided assistance on tools for initializing MPAS. We also thank Hugh Morrison (NCAR) for valuable discussions and insights related to data analysis. A. Roseman joined under the NSF NCAR Advanced Study Program (ASP) Graduate Visitor Program (GVP). A. Roseman and G. Torri are partially supported by the Department of Energy award number DE-SC0024265 and National Science Foundation award number OIA-2327232.

%
%
\datastatement

MPAS model meshes are publicly available from the official MPAS website (\url{https://mpas-dev.github.io/}). Model initial and boundary conditions are derived from the ERA5 reanalysis, available through the Copernicus Climate Data Store (\url{https://cds.climate.copernicus.eu/}). Simulations in this study can be reproduced using the ERA5 dataset and MPAS meshes. NSSL and TEMPO microphysics for MPAS are located in the GSL version of MPAS (\url{https://github.com/MicroTed/MPAS-Model.git} or \url{https://github.com/ufs-community/MPAS-Model}). Observational data used in this study include the Multi-Radar/Multi-Sensor (MRMS) system (\url{https://mrms.ncep.noaa.gov/data/}) and data from the PRECIP 2022 field campaign (\url{http://precip.org/}), which are available upon request and subject to data access permissions. Data from the 2021-2022 Tracking Aerosol Convection Interactions Experiment (TRACER) campaign are available from Atmospheric Radiation Measurement (ARM) (\url{https://armgov.svcs.arm.gov/research/campaigns/amf2021tracer}). Hawai\okina i weather station date availability from University of Utah MesoWest (\url{https://mesowest.utah.edu/}). Area Forecast Discussions are available from Iowa Environmental Mesonet (\url{https://mesonet.agron.iastate.edu/}).

%






%




\bibliographystyle{ametsocV6}
\bibliography{references}

\end{document}